\documentclass[lettersize,journal]{IEEEtran} 
\usepackage{amsmath,amsfonts}
\usepackage{algorithmic}
\usepackage{algorithm}
\usepackage{array}
\usepackage[caption=false,font=normalsize,labelfont=sf,textfont=sf]{subfig}
\usepackage{textcomp}
\usepackage{stfloats}
\usepackage{url}
\usepackage{verbatim}
\usepackage{graphicx}
\usepackage{cite}
\usepackage[switch]{lineno} 
\usepackage{color}
\usepackage{enumerate}

\usepackage{fancyhdr}
\begin{document}

\title{Integration of Independent Heat Transfer Mechanisms for Non-Contact Cold Sensation Presentation with Low Residual Heat}

\author{Jiayi~Xu,~\IEEEmembership{\textcolor{black}{Member}
,~IEEE},
Shoichi Hasegawa,~\IEEEmembership{Member,~IEEE}, 
Kiyoshi Kiyokawa,~\IEEEmembership{Member,~IEEE}, Naoto~Ienaga,~\IEEEmembership{Member,~IEEE}, and~Yoshihiro~Kuroda,~\IEEEmembership{Member,~IEEE}
\thanks{Manuscript received April xx, xxxx; revised August xx, xxxx. This work was supported in part by grants from JSPS KAKENHI (JP21H03474, JP21K19778) and in part by JST SPRING (JPMJSP2124).}
\thanks{J. Xu is with the \textcolor{black}{Institute of Systems and Information Engineering,}
University of Tsukuba, Japan (e-mail: \textcolor{black}{xu.jiayi@ms.esys.tsukuba.ac.jp}).}

\thanks{S. Hasegawa is with the Institute of Innovative Research, Tokyo Institute of Technology, Japan.}
\thanks{K. Kiyokawa is with the Graduate School of Science and Technology, Nara Institute of Science and Technology, Ikoma, Japan.}
\thanks{N. Ienaga and Y. Kuroda are 
with the \textcolor{black}{Institute of Systems and Information Engineering}, University of Tsukuba, Japan.}}

\maketitle 
\thispagestyle{fancy}
\begin{abstract}
\textcolor{black}{
Thermal sensation is crucial to enhancing our comprehension of the world and enhancing our ability to interact with it. Therefore, the development of thermal sensation presentation technologies holds significant potential, providing a novel method of interaction. Traditional technologies often leave residual heat in the system or the skin, affecting subsequent presentations. Our study focuses on presenting thermal sensations with low residual heat, especially cold sensations. To mitigate the impact of residual heat in the presentation system, we opted for a non-contact method, and to address the influence of residual heat on the skin, we present thermal sensations without significantly altering skin temperature. Specifically, we integrated two highly responsive and independent heat transfer mechanisms: convection via cold air and radiation via visible light, providing non-contact thermal stimuli. By rapidly alternating between perceptible decreases and imperceptible increases in temperature on the same skin area, we maintained near-constant skin temperature while presenting continuous cold sensations. In our experiments involving 15 participants, we observed that when the cooling rate was $-$0.2 to $-$0.24~$^\circ$C/s and the cooling time ratio was 30 to 50\%, more than 86.67\% of the participants perceived only persistent cold without any warmth.}
\end{abstract}


\section{Introduction}
\IEEEPARstart{I}{n} our daily life, thermal sensation plays an important role in enriching our perception of the world and facilitating our interaction with it\cite{warmAndCool}. With the advancement of virtual reality (VR) technology and the emergence of Metaverse, the demand for thermal sensation presentation technologies is increasing in multisensory interactive systems. With these technologies, we can create a novel method of interaction that allows users to interact with virtual worlds using their bare skin through the transfer of heat. For example, we can represent being close to a campfire while camping, cold winds blowing across cheeks while skiing, 
\textcolor{black}{and other interactions that involve thermal sensations. To achieve these interactions naturally, it is essential that a method can stably present accurate thermal sensations according to a user’s input.}
\begin{figure}[t]
	\centering
	\subfloat[]{\includegraphics[scale=0.33]{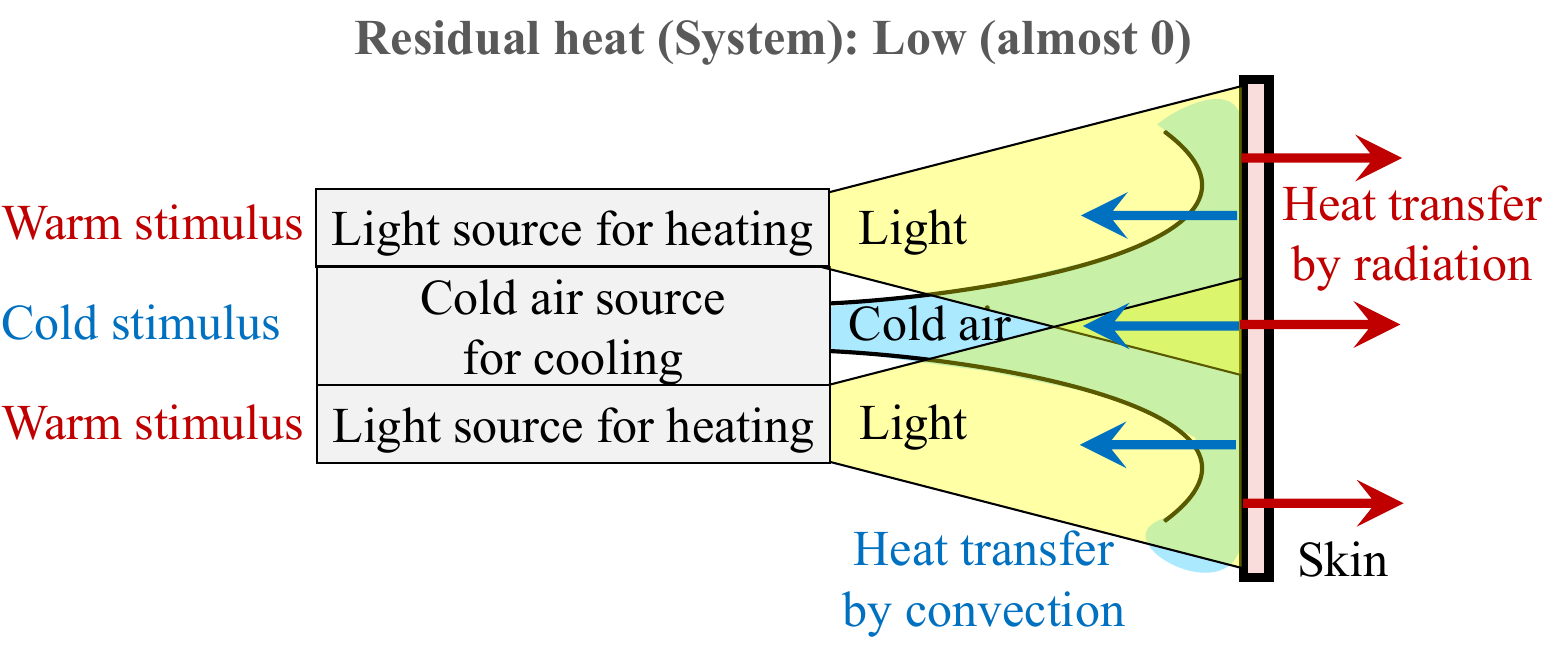}}
    \hfil
    \subfloat[]{\includegraphics[scale=0.33]{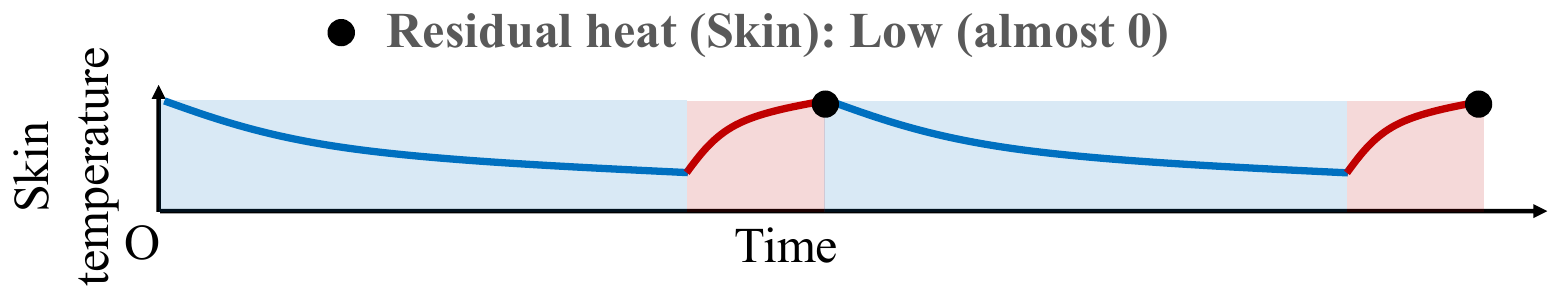}}
    \hfil
    \subfloat[]{\includegraphics[scale=0.33]{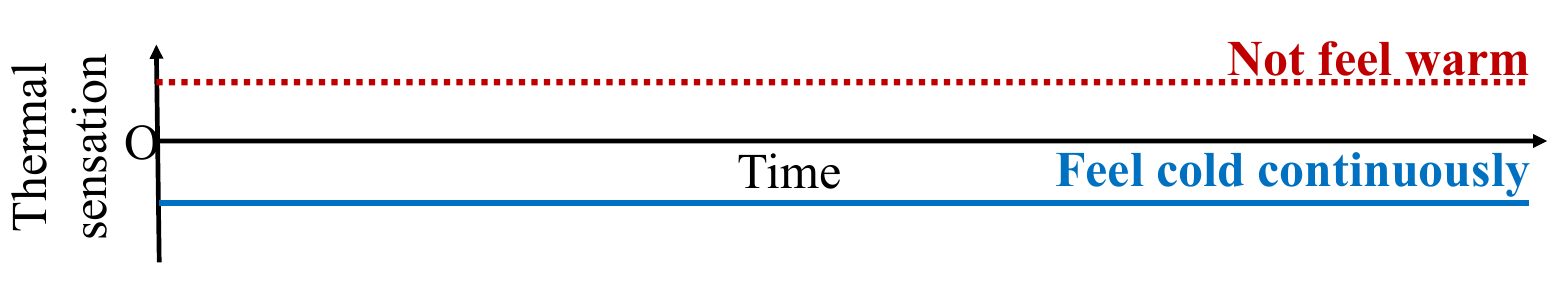}}
	\caption{Overview of the proposed method. \textcolor{black}{(a) We utilize the convection heat transfer mechanism for providing cold stimuli and the radiation heat transfer mechanism for providing warm stimuli in a non-contact manner to mitigate the influence of residual heat in the presentation system. (b) We alternately change the temperature rise and fall in a non-contact manner by providing cold and warm stimuli in the same skin area to ensure that the residual heat in the skin per stimulation cycle is kept to almost zero. (c) With the design of the stimulus pattern, it is possible to make users feel only persistent cold without any warmth.}
 }	
 \label{fig:Overview}
\end{figure}

\par \textcolor{black}{To stably and accurately present thermal sensations, managing the heat remaining both in the skin and the presentation system after thermal stimulation is crucial. The heat remaining in the skin could prompt changes in skin temperature, thus affecting thermal sensitivity~\cite{temperatureSensitivity}. For instance, when the skin temperature is low, small temperature drops can elicit cold sensations, whereas significant temperature rises are required to elicit warm sensations; when the skin temperature is high, the opposite occurs. Consequently, the complexity of achieving the targeted thermal sensation escalates when skin temperature changes. On the other hand, the heat remaining in the system could potentially induce unnecessary additional stimuli during non-stimulation phases. We refer to the heat remaining both in the skin and the presentation system, which can influence subsequent thermal sensation presentations, as {\it residual heat}.}

\par \textcolor{black}{Conventional thermal sensation presentation technologies elicited thermal sensations by continuously raising or falling skin temperature~\cite{peltier1,peltier2,peltier3,peltier4,peltier5,peltier6,peltier7,water1,water2,fan,fan1,dryIce,coldAir,coldAir2,coldAir3,hot1,hot2,hot3,hot4,hot5,coldRadient,hokoyama2017mugginess,mist0,mist1,mist2}. For instance, some studies utilized electromagnetic waves to warm the skin and presented warm sensations by continuously increasing the skin temperature~\cite{hot1,hot2,hot3,hot4,hot5}. Additionally, we developed a method that employed cold air to induce a sense of coldness by continuously lowering the skin temperature in our previous study~\cite{coldAir,coldAir2,coldAir3}. These approaches are closer to the natural experience, such as the continuous gain of heat when we put our hand near the fire or
the continuous loss of heat when the snow blows across our cheeks. However, they faced the issue of residual heat on the skin, and it could be inconvenient for VR systems to simulate different temperature experiences within a short time. }
\par \textcolor{black}{To address the issue of residual heat in the skin, the method proposed by Manasrah et al., which elicits thermal sensations without changing the average temperature of the skin, offers valuable insights~\cite{continuousThermal1,continuousThermal2,continuousThermal3}. They arranged a grid of separately controlled Peltier elements, allowing for an alternation of skin temperature between cooling and warming across different skin areas. For example, when presenting a cold sensation, some elements exhibit noticeable temperature decreases, while others display imperceptible temperature rises. The continuous cold sensation is achieved by ensuring that there are always some elements in a state of perceptible temperature decrease. It should be noted that Peltier elements facilitate the creation of disparate thermal interfaces — hot and cold sides — via the process of heat transfer. Nonetheless, this process concurrently results in an aggregate heat production within the entire system. Given that they necessitate direct contact with the skin, if heat dissipation is insufficient, the residual heat in the presentation system may transfer to the skin, potentially affecting the presentation of subsequent sensations.}

\par \textcolor{black}{In this study, we focus on cold sensations and aim to develop a method that presents thermal sensations with low residual heat. To mitigate the influence of residual heat in the presentation system, we propose adopting a non-contact approach, thereby avoiding the transfer of residual heat from the presentation component to the skin through direct contact. To effectively reduce the influence of residual heat in the skin, we draw inspiration from the work of Manasrah et al. and aim to develop a method that can elicit cold sensations while maintaining a nearly constant skin temperature.}
\par \textcolor{black}{Compared to contact-type methods, achieving precise and strictly controlled spatial distribution of thermal stimuli through non-contact methods is challenging. Therefore, we contemplated an alternative approach where skin temperature alternates within the same area. 
We hypothesize that {\ even if there is a period where the skin temperature does not decrease after applying cold stimuli, people perceive continuous cold}. This hypothesis is based on an everyday optical phenomenon: when a light rapidly alternates between on and off states, we perceive continuous brightness. By rapidly alternating perceptible temperature decrease (similar to a light on) and imperceptible temperature increase (similar to a light off) within the same skin area, it is possible to elicit continuous cold sensations while keeping the skin temperature almost unchanged.}
\par \textcolor{black}{Specifically, to mitigate the influence of residual heat in the presentation system, we employed a combination of a cold air source utilizing convection heat transfer and a light source utilizing radiation heat transfer to provide cold and warm stimuli, respectively, in a non-contact manner, as shown in Fig.~\ref{fig:Overview}~(a). Because these mechanisms are independent, they can be integrated to synthesize the skin temperature changes. We alternately changed the temperature rise and fall in a non-contact manner by providing cold and warm stimuli in the same skin area. The residual heat in the skin was maintained at almost zero per stimulation cycle owing to a dynamic balance between the amount of heat input and heat output, as shown in Fig. ~\ref{fig:Overview}~(b). The high responsiveness of our non-contact cooling and warming methods allows the stimulation cycle to be minimized to 0.5 s, making it possible to achieve persistent cold sensations, as shown in Fig. ~\ref{fig:Overview}(c).}
\par In this study, we reported the proposed method (refer to Section \uppercase\expandafter{\romannumeral3}) and constructed a prototype system (refer to Section \uppercase\expandafter{\romannumeral4}). Moreover, we conducted experiments to investigate the responses of the proposed method to human perception (refer to Section \uppercase\expandafter{\romannumeral5}). The experimental results allow us to 
\textcolor{black}{confirm our hypothesis, }
determine the cold sensations that our system can provide, as well as the conditions for eliciting non-contact cold sensations with low residual heat.

\section{Related Works}
\par In this section, we first introduce the perceptual characteristics of thermal sensations. We then introduce the heat transfer mechanisms and review related works on eliciting thermal sensations based on these mechanisms. Finally, we describe the position of our study.
\subsection{Perceptual Characteristics of Thermal Sensations}
\par There are cold and warm receptors in the skin that can encode a wide range of temperatures. People are more sensitive to cold partly because they have more cold receptors than hot receptors. Cold receptors are maximally responsive between 22 and 28 $^\circ$C, whereas warm receptors are maximally responsive at approximately 45 $^\circ$C~\cite{thermoreceptors,temperatureReceptors}. Therefore, thermal sensitivity is closely related to changes in skin temperature\cite{temperatureSensitivity}. When the skin temperature is low, only small temperature drops are required to elicit cold sensations, whereas large temperature changes are required to elicit warm sensations. In contrast, when the skin temperature is high, only small temperature rises are required to elicit warm sensations, whereas larger temperature changes are required to elicit cold sensations. That is, thermal sensitivity is not constant but varies depending on the temperature of the skin.
\par In addition, the speed of temperature change affects temperature perception\cite{thresholds}. When the skin temperature changes rapidly, humans can even detect a slight temperature change; however, if the skin temperature changes slowly, the thermal sensation is not felt until there is a significant temperature change. \textcolor{black}{Manasrah \textit{et al.}~\cite{continuousThermal1,continuousThermal2,continuousThermal3} utilized this phenomenon and proposed a method to elicit thermal sensations while maintaining skin temperature almost unchanged. They found that people perceived only coldness without any warmth when perceptible temperature decreases alternated with imperceptible temperature increases in several areas of the skin, and vice versa.}
\subsection{Eliciting Thermal Sensations}
\par Skin can exchange heat with the environment through four heat transfer mechanisms~\cite{heattransfer1}\cite{heattransfer2}: conduction, convection, radiation, and evapotranspiration. Owing to the importance of integrating different mechanisms in this study, we describe the methods that utilize each mechanism.
\subsubsection{Methods based on conduction heat transfer}
\par Conduction heat transfer occurs when the skin is in contact with an object of a different temperature. The heat is transferred from the higher-temperature area to the lower-temperature area. Thermal displays based on conduction heat transfer are always of the contact type. For example, Peltier devices are used to construct contact thermal displays because they are small and easy to assemble\cite{peltier1,peltier2,peltier3,peltier4,peltier5,peltier6,peltier7,continuousThermal1,continuousThermal2,continuousThermal3}. Sakaguchi \textit{et al.}~proposed a display using water as the heat source\cite{water1,water2}. Rapid temperature changes can be achieved by switching the cold or warm water transport component to the presentation component. However, we did not use this method because we wished to elicit cold sensations without physical contact.
\subsubsection{Methods based on convection heat transfer}
\par Convection heat transfer occurs by the movement of fluids. Natural convection occurs because of the temperature difference between the skin and air. Conversely, forced convection occurs when heat transfer is caused by a mechanical method, such as an electric fan. An air conditioner is a traditional way to present thermal sensations based on convection. However, it takes time to change the room temperature by replacing a large amount of air; thus, it is impossible to switch thermal sensations according to user input in real time. In addition, some studies proposed methods that used fans, blowers, or cold air generated with dry ice to elicit non-contact cold sensations~\cite{fan,fan1,dryIce}. Nevertheless, these methods have various shortcomings, including the inability to significantly reduce the skin temperature because the air must be precooled for a long time. In previous studies, we proposed a cooling method based on the vortex effect\cite{coldAir,coldAir2,coldAir3}. \textcolor{black}{The vortex effect is the phenomenon of thermal separation occurring when air swirls within a tube. This effect was discovered in thermal engineering and is generated by compressed air injected into a tube, causing it to swirl and separate into cold and warm air.} Because the vortex effect can generate low-temperature cold air immediately from compressed air, the generated cold air can be transferred to quickly reduce the skin temperature and elicit strong cold sensations. Additionally, changing the volume flow rate of cold air allows for easy adjustment of sensation intensity. \textcolor{black}{The volume flow rate of cold air at different duty cycles can be found in our previous study~\cite{coldAir2}.}
\subsubsection{Methods based on radiation heat transfer}
\par Radiation heat transfer occurs when the skin absorbs or emits electromagnetic waves. Lasers, infrared lamps, and halogens are used as heat sources for non-contact warm sensation displays\cite{hot1,hot2,hot3,hot4}. Owing to their high temperatures, they emit electromagnetic waves that warm the skin. Consequently, when these heat sources were turned off, excessive heat was retained. Sakai~\textit{et al.} proposed a method that utilizes light-emitting diodes (LEDs) to warm skin\cite{hot5}. LEDs emit electromagnetic waves in the form of visible light from semiconductor PN junctions regardless of their temperature. Because LEDs do not transmit electromagnetic waves through their heated bodies, they do not retain as much heat as other heat sources when turned off. Consequently, the LEDs can be quickly switched on and off to warm the skin. 
Additionally, Kume~\textit{et al.} proposed a method for eliciting cold sensations using a cold insulating material as a cryogenic radiation source\cite{coldRadient}. Nevertheless, it requires shutting out any other heat sources around the user. 
\subsubsection{Methods based on evapotranspiration heat transfer}
\par Evaporative heat transfer occurs when moisture in the skin undergoes a phase change, such as evaporation or sublimation. There are several methods based on evapotranspiration heat transfer for presenting thermal sensations. For example, Hokoyama~\textit{et al.} proposed a system for presenting a warm sensation by increasing the humidity of the air around the skin\cite{hokoyama2017mugginess}. In addition, Nakajima~\textit{et al.} used ultrasonic waves to deliver water mist to the skin surface to provide a sudden pinpoint cooling sensation\cite{mist0,mist1,mist2}. 

\subsection{Positioning of This Study}
\par \textcolor{black}{Our research hypothesizes that {\it even when the skin temperature does not decrease for a period after cold stimuli have been provided, people perceive continuous coldness.} Based on this hypothesis, we aim to develop a method that, by rapidly alternating perceptible temperature decreases and imperceptible temperature increases in the same skin area, enables individuals to perceive coldness continuously without significant changes to their skin temperature. Since people are more sensitive to cold than to warmth, this approach is more likely to realize the presentation of cold sensations compared to warm sensations.} In this study, the focus was on non-contact cold sensations, which indicates the necessity for non-contact methods capable of providing rapid temperature changes.
\textcolor{black}{Using convection, radiation, and evapotranspiration heat transfer mechanisms, cold and warm stimuli can be delivered without direct contact.} 
Among these mechanisms, evapotranspiration alters the skin moisture, which may complicate the presentation of subsequent thermal sensations. 
\textcolor{black}{Therefore, we did not utilize the mechanism based on evapotranspiration heat transfer.}
Using radiation heat transfer is a simple process to present warm stimuli, but presenting cold stimuli is challenging. Convection heat transfer involves both hot and cold air, providing warm and cold stimuli; however, these stimuli interfere with each other.
\textcolor{black}{For this end, we combined radiation heat transfer for heating with convection heat transfer for cooling to develop a method that can elicit non-contact cold sensations with low residual heat. Specifically, we chose to present cold stimuli using cold air generated with the vortex effect, which is one of the most effective ways to convey cold sensations through convection heat transfer. For warm stimuli, we chose to use LEDs that emit electromagnetic waves, namely light, directly without needing a heated surface. This enables us to instantly switch on/off warm stimuli based on radiation heat transfer.}

\section{Method}
\par In this section, we detail our proposed method to elicit cold sensations with low residual heat in a non-contact manner. Specifically, we combined a cold air source that utilized convection heat transfer for cooling and a light source that utilized radiation heat transfer for warming, as shown in Fig. ~\ref{fig:Overview} (a). Because these mechanisms are independent, they can be integrated to synthesize the skin temperature changes. As shown in Fig.~\ref{fig:proposedMethod}, we alternately provided cold stimuli continuously and warm stimuli intermittently to alternately change the temperature rise and fall in a non-contact manner, as cold air cannot be turned on or off as quickly as LEDs. The residual heat per stimulation cycle was maintained at almost zero owing to the dynamic balance between the amount of heat input and heat output. Designing the stimulus pattern is expected to create a persistent cold sensation without any warmth while maintaining an almost constant skin temperature.

\subsection{Design of the Stimulus Pattern.}
\par
\begin{figure}[t]
    \centering
    \includegraphics[scale=0.33]{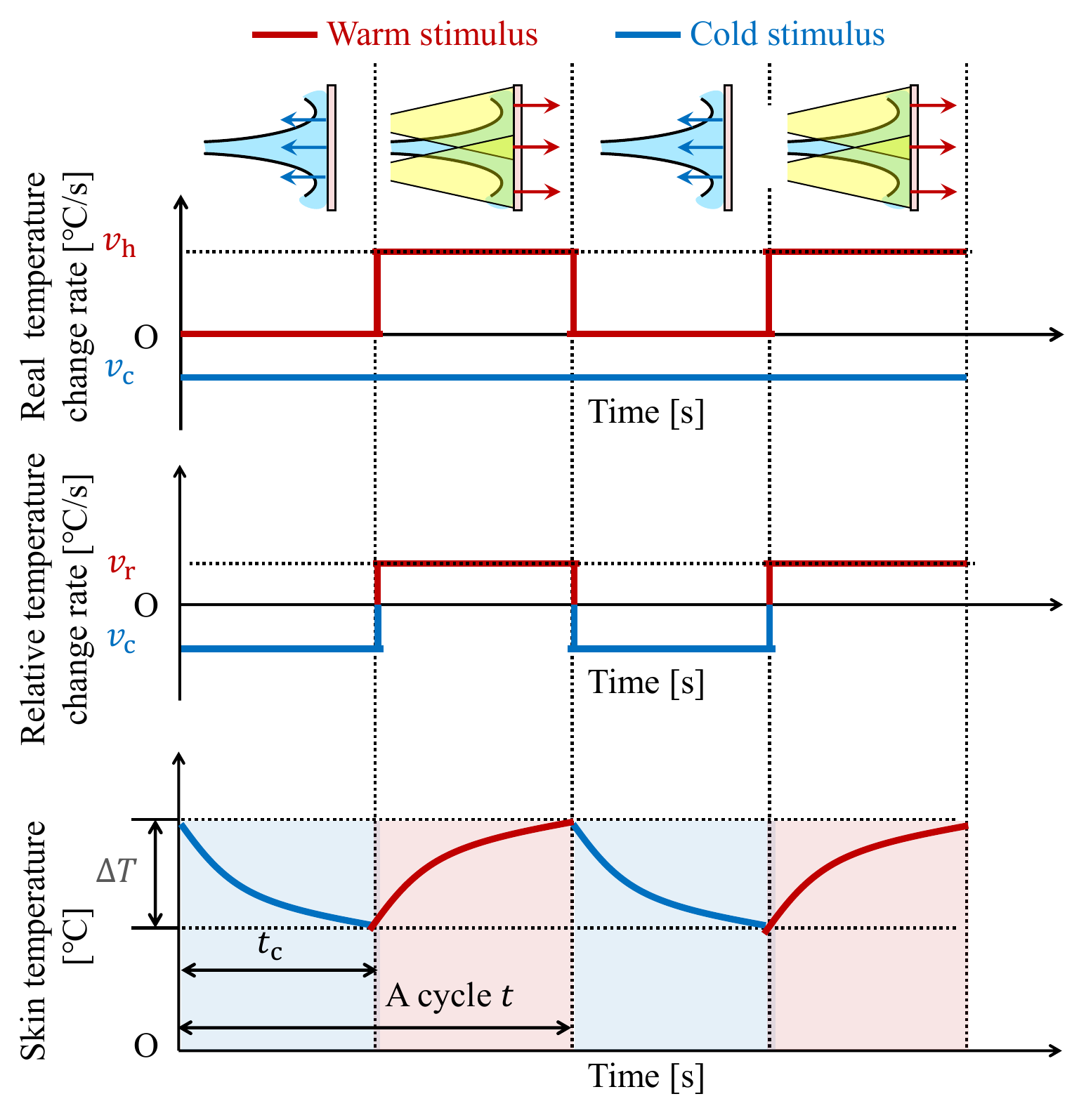}
    \caption{Design of the stimulus pattern. \textcolor{black}{We alternately provided continuous cold and intermittent warm stimuli to induce alternating temperature rises and falls in a non-contact manner. This approach was chosen as cold air cannot be turned on or off as quickly as LEDs. Skin temperature changes were calculated based on synthesized temperature change rates.}}
    \label{fig:proposedMethod}
\end{figure}

 As shown in Fig.~\ref{fig:proposedMethod}, in a cycle with duration $t$, during the cooling period $t_{\rm c}(<t)$, only the cold stimulus is presented, while during the warming period $t-t_{\rm c}$, both cold and warm stimuli are presented simultaneously to restore the skin temperature. During the cooling and warming periods, the skin temperature decreased and then increased by $\Delta T(>0)$. 
We defined the cooling rate $v_{\rm c}$ as the average rate of temperature change during period  $t_{\rm c}$; therefore,
\begin{equation}
      v_{\rm c}=-\frac{\Delta T}{t_{\rm c}}.
\end{equation}
The cooling time ratio $\lambda_{\rm c}$ is defined as the ratio of the cooling time in one cycle:
\begin{equation}
      \lambda_{\rm c}=\frac{t_{\rm c}}{t}.
\end{equation}
During the warming period $t-t_{\rm c}$, the relative warming rate $v_{\rm r}$, which is the addition of cooling and warming effects to restore skin temperature, is represented as
\begin{equation}
      v_{\rm r}=\frac{\Delta T}{t-t_{\rm c}}
      =\frac{- t_{\rm c}}{t-t_{\rm c}}v_{\rm c}
      =\frac{- \lambda_{\rm c}}{1-\lambda_{\rm c}}v_{\rm c}.
\end{equation}
During that period, because the cold and warm stimuli are presented simultaneously, the warming rate $v_{\rm h}$ that we need to give can be calculated as
\begin{align}
     v_{\rm h}&= v_{\rm r}-v_{\rm c}\nonumber\\
      &=-\dfrac{v_{\rm c}}{1-\lambda_{\rm c}}
\end{align}
Once we determined the cooling rate $v_{\rm c}$, cooling time ratio $\lambda_{\rm c}$, and temperature change $\Delta T$, the other parameters were also determined. For example, if we design a stimulus with a cooling rate $v_{\rm c}$ of $-0.1$~$^\circ$C/s, a cooling time ratio $\lambda_{\rm c}$ of 50~\%, and a temperature change $\Delta T$ of 0.06~$^\circ$C, then the cooling time $t_{\rm c}$ should be 0.6 s, total time of one cycle $t$ should be 1.2 s, and warming rate $v_{\rm h}$ should be 0.2~$^\circ$C/s.
\section{System}
\par In this section, we detail the system design and implementation for the method to elicit cold sensations with low residual heat.
\subsection{Cold Air Source for Cooling}
\par We chose to present cold stimuli using a low-temperature cold air source that applies the vortex effect discovered in thermal engineering, one of the most effective ways to present cold sensations based on convection heat transfer in a non-contact manner. A cold air source for cooling is a device that can provide low-temperature cold air instantly from a compressed air supply~\cite{vortextube}. In addition, the intensity of the cold sensations can be easily adjusted by adjusting the volume flow rate of the generated cold air.

\subsection{Light Source for Warming}
\par We chose to present warm stimuli using LEDs, which emit light directly without a heated body and can immediately switch on/off warm stimuli based on radiation heat transfer in a non-contact manner. The skin absorbs the light emitted by the LED, and the skin temperature increases. In addition, the intensity of warm sensations can be easily adjusted based on the brightness of light.

\subsection{System Configuration}
\par
\begin{figure}[b]
    \centering
    \subfloat[]{\includegraphics[scale=0.31]{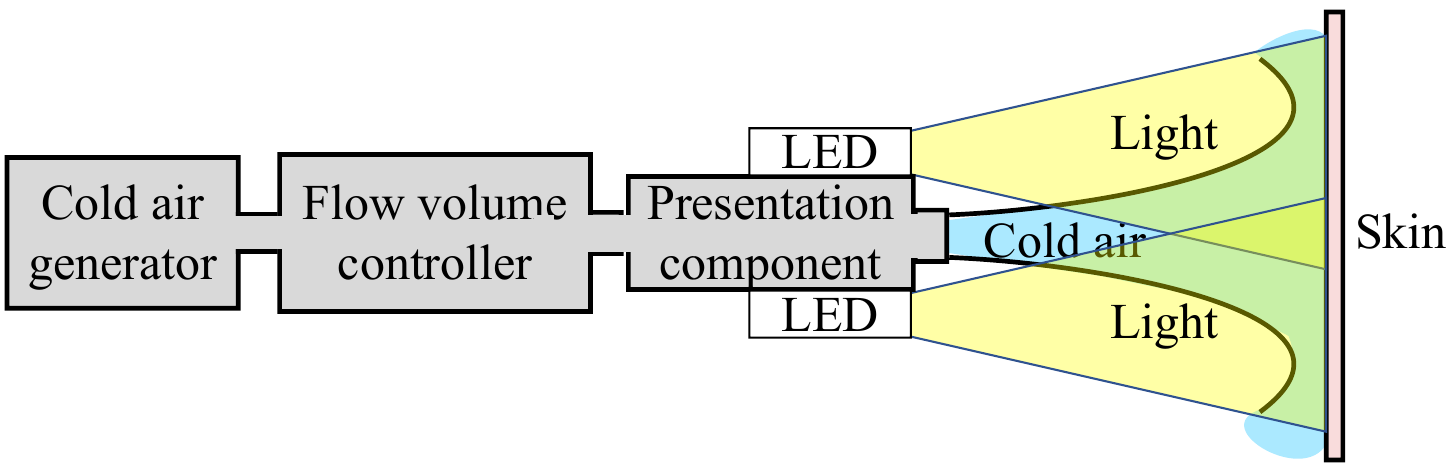}}\\
    \subfloat[]{\includegraphics[scale=0.31]{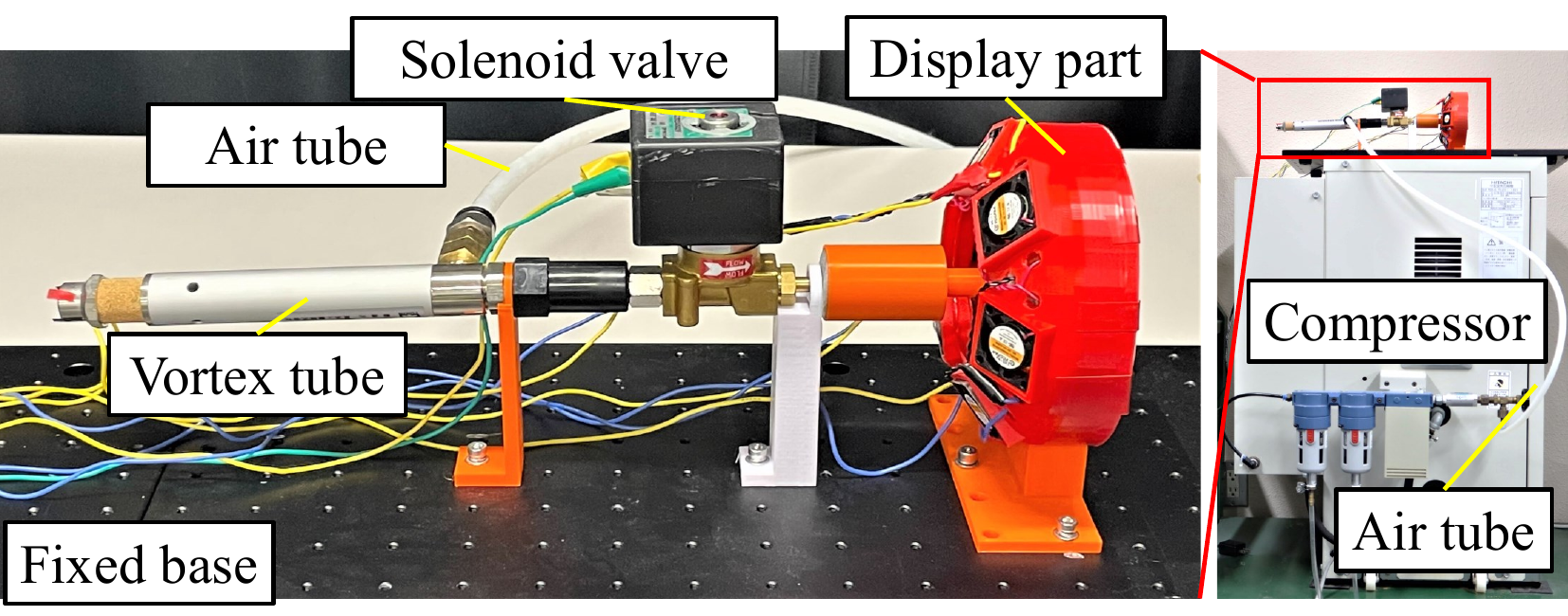}}
    \caption{\textcolor{black}{System design and implementation: (a) Configuration of the proposed system. (b) Implementation of the proposed system.} 
    }
    \label{fig:System}
\end{figure}

 Fig.~\ref{fig:System}~(a) shows the configuration of the proposed system. We adjusted the flow volume of the generated cold air to present the cold stimulus, and the brightness of the LEDs to present the warm stimulus. The combination of the two stimuli resulted in the target temperature changes. The prototype system consists of a cold air generator, flow volume controller, and presentation component. 
As shown in Fig.~\ref{fig:System}~(b), each was mounted using a vortex tube (Tohin/AC-50), solenoid valve (Asco/Positive-flow-202), and presentation component printed by a 3D printer (Raise3D/Pro2 with PLA filament). A compressor (Hitachi/POD-0.75LES) was used to supply compressed air to the vortex tube. The generated air pressure by the compressor is maintained at approximately 0.6 MPa to keep the temperature of the generated cold air at approximately 0 $^\circ$C. In addition, it is possible to mechanically adjust the cold air ratio of the vortex tube, which is the volume of the output cold air relative to the input compressed air. When the cold air ratio was high, the amount of output cold air was large, whereas the temperature difference between the input and output was slight. In this system, we set the cold-air ratio to 75\%.

\subsection{Presentation Component}
\par
\begin{figure}[t]
    \centering
    \subfloat[]{\includegraphics[scale=0.3]{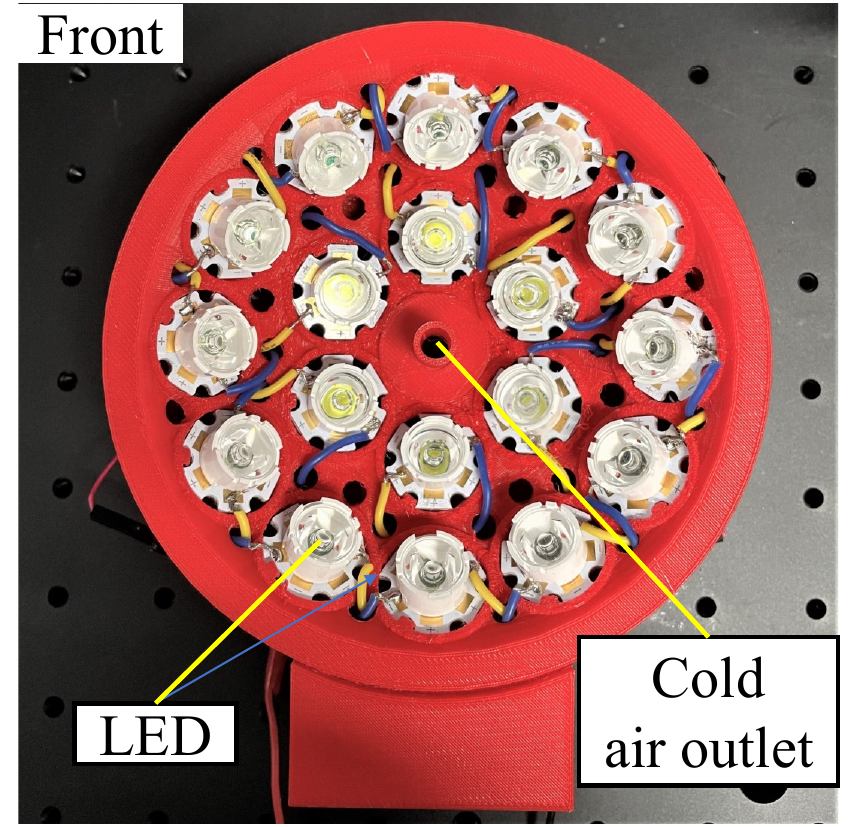}}   
    \subfloat[]{\includegraphics[scale=0.3]{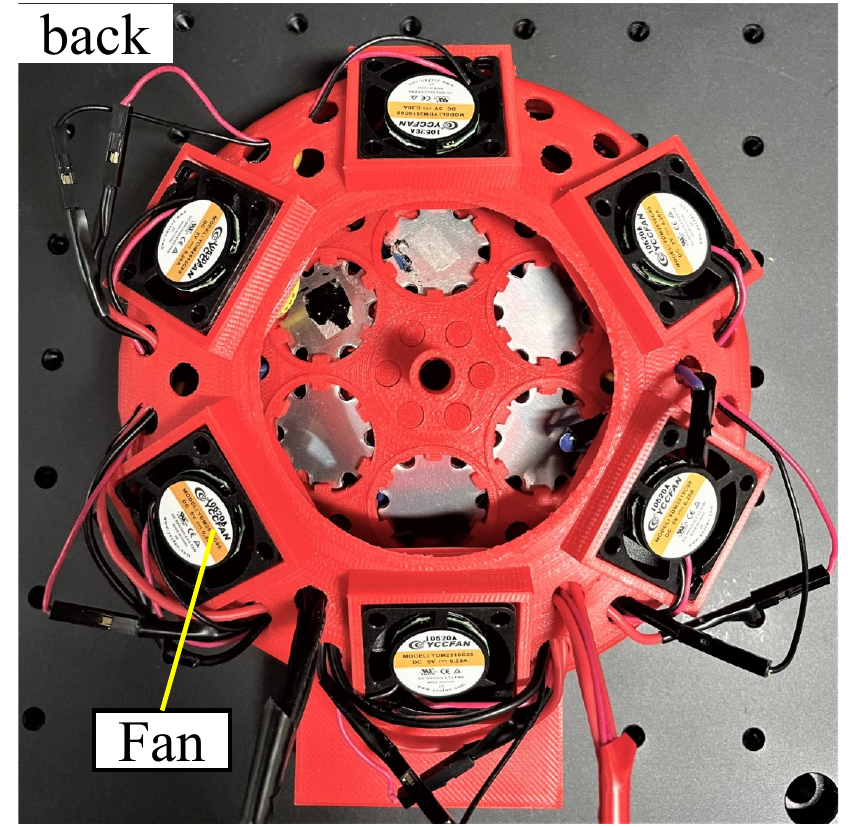}}
    \caption{\textcolor{black}{Presentation component: (a) Front side. (b) Back side.}
    }
    \label{fig:Presentation component}
	\centering
	\includegraphics[scale=0.45]{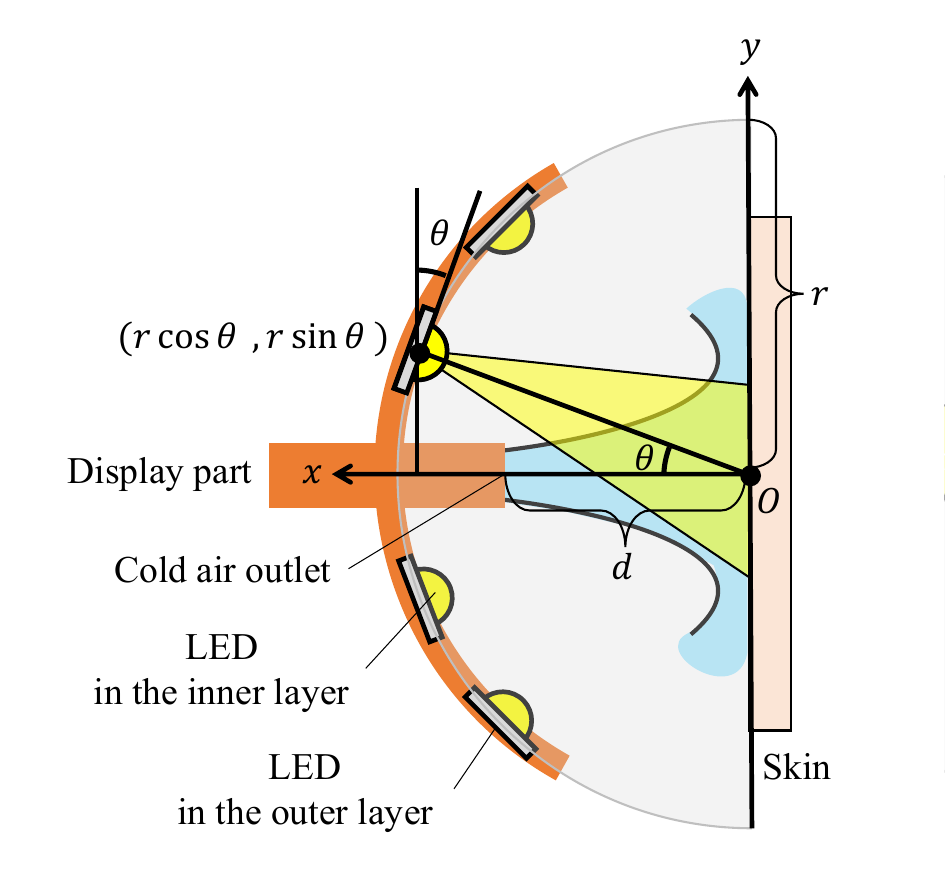}
	\caption{Design of the presentation component.}
	\label{fig:Design of the presentation component}
\end{figure}

 As shown in Fig.~\ref{fig:Presentation component}, we placed a cold air outlet and 18 LEDs (CREE/XPGWHT-L1-STAR-G53 5~W) with heat-dissipation boards on the presentation component. 
We attached a lens (CREE/LL01CR-DF40L06-M2-T FWHM:28°) to each LED to collect light, and six fans (Nidec/D02X-05TS1 02 5V DC) to the heat dissipation board for proper heat dissipation. To present cold and warm stimuli in the same skin area, we needed to make the temperature distribution due to the warm and cold stimuli as consistent as possible. As shown in Fig.~\ref{fig:Design of the presentation component}, we drew a hemisphere of radius $r$ with the center $O$ of the skin surface as the origin (we drew in 2D for simplicity). The chief ray of the LED light perpendicular to the hemisphere directly hits the center $O$ of the skin surface; that is, the stimulus given to the center of the skin surface is the strongest, co-located with the cold stimulus. As long as the LED rotation angle $\theta$ is determined, we can calculate where the LED should be. For example, we established a rectangular coordinate system XY, as shown in Fig. ~\ref{fig:Design of the presentation component}. The LED position is ($r\cos{\theta}$ and $r\sin{\theta}$). In our design, we set the radius of the hemisphere to 60 mm. The LEDs are distributed in two layers, as shown in Fig. ~\ref{fig:Presentation component}(a). \textcolor{black}{The rotation angle of the 6 LEDs in the inner layer was 20.5°, and that of the 12 LEDs in the outer layer was 45°.} The distance $d$ from the cold air outlet to the skin surface was 42 mm. \textcolor{black}{The selection of a presentation distance of 42 mm is based on the heat transfer characteristics of the jet flow. Previous research has demonstrated that the heat transfer characteristics are optimal when the presentation distance is 6-8 times the diameter of the flow outlet~\cite{jetEngineering}. Therefore, we chose a presentation distance of 42 mm, which is 7 times the diameter of the cold air outlet (6 mm), to ensure the ideal heat transfer conditions.}

\subsection{System Control}
\par We controlled the output flow volume of the solenoid valve and the output brightness of the LEDs by the pulse width modulation (PWM) method through serial communication with the microcomputer (Arduino/Due). We used a high-power MOSFET trigger switch drive module (Baoblaze/B07BVKL4VW) for PWM control of the solenoid valve. We used two LED drive module  (STMicroelectronics/STCS2SPR) for PWM control of the LEDs.

\section{User study}
\par \textcolor{black}{To evaluate the proposed system and method, we conducted three experiments with participants. The primary objective of these experiments was to explore how participants perceive and evaluate the persistence and intensity levels of cold sensation when exposed to various stimulus patterns. Here is a breakdown of each experiment:
\begin{itemize}
    \item The first experiment aimed to measure and calibrate the system output. This process involved determining the duty ratio of the PWM corresponding to the required cooling and warming rates necessary for implementing specified temperature change patterns, that is, the stimulus patterns. 
    Subsequently, we verified whether the skin temperature remained almost unchanged before and after the application of these specific stimulus patterns. If there were obvious changes in skin temperature, we implemented additional calibration to the relationship between the duty ratio and the warming rate until we achieved a state where the skin temperature remained almost unchanged before and after the application of the specified stimulus patterns. This measurement and calibration process was essential in preparing for the subsequent experiments, as it ensured that the system output could accurately produce the intended stimulus patterns.
    \item The second experiment aimed to investigate the persistence of the cold sensation under different stimulus patterns. Participants were exposed to various stimulus patterns, and their assessment of thermal sensations was recorded. The primary objective of this experiment was to gain insights into the specific stimulation conditions that can induce a persistent sensation of coldness, and to discuss the hypotheses we formulated.
    \item The third experiment aimed to assess the intensity of the cold sensation generated by different stimulus patterns. Participants were exposed to various stimulus patterns, and they were asked to rate the intensity of the cold sensation they experienced. This experiment aimed to understand how different patterns of stimulation influenced the perceived intensity of the cold sensation.
\end{itemize}
}

\subsection{Participants}
\par A total of 15 paid participants (aged 20--30 years, 1 female and 14 males) were enrolled. Each participant received approximately USD 11 (1,500 JPY) in the form of an Amazon gift card as monetary compensation. The recruitment of participants and experimental procedures were approved by the Ethical Committee of the Faculty of Engineering, Information, and Systems, University of Tsukuba, Japan (approval number 2020R453). All participants provided written informed consent before participating in the study.

\subsection{Experimental Settings}
\par
\begin{figure}[b]
\centering\includegraphics[scale=0.28]{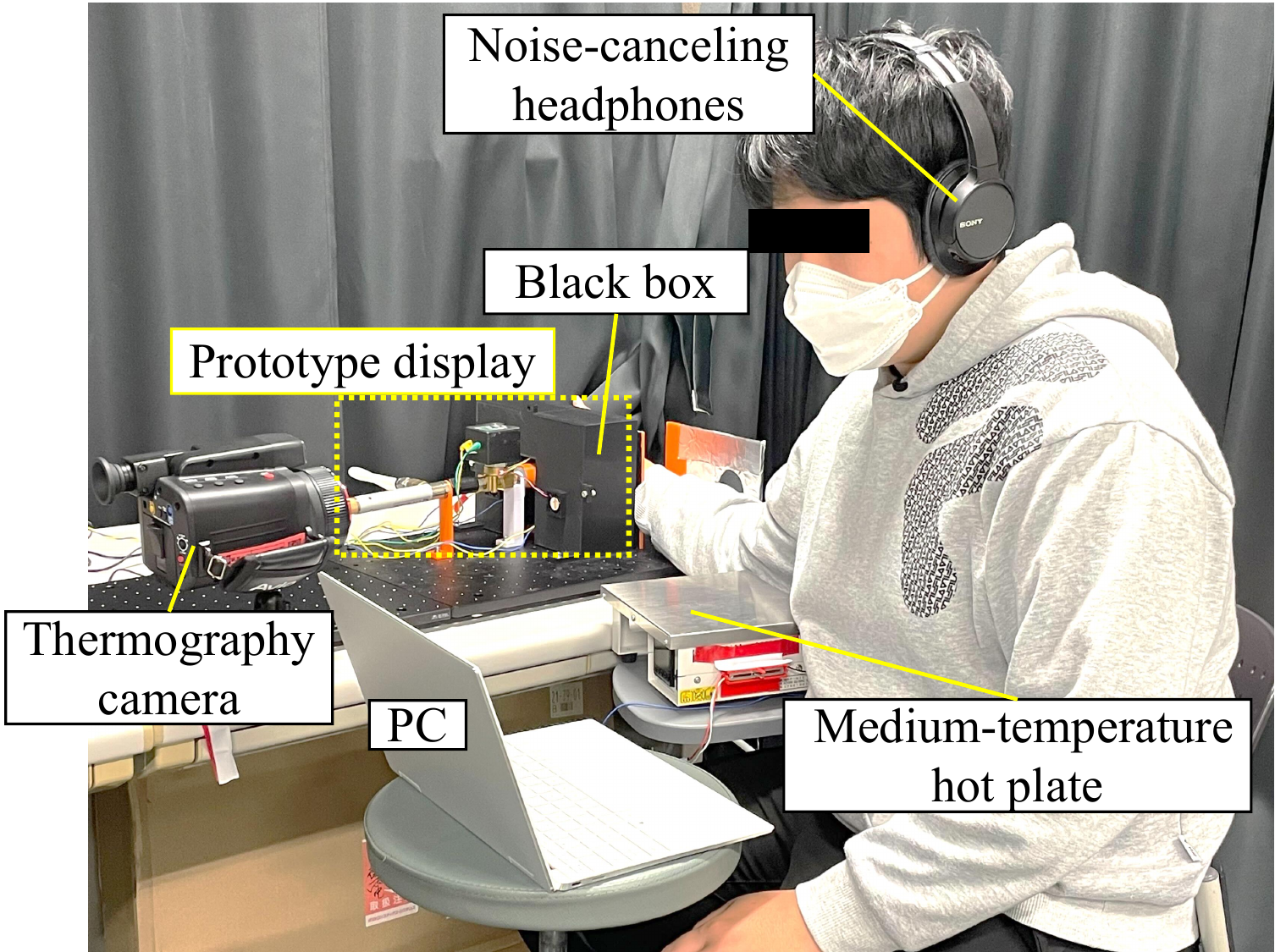}
	\caption{Experimental settings.}
	\label{fig:Experimental settings}
\end{figure}

 As shown in Fig.~\ref{fig:Experimental settings}, we covered the prototype display with a black box to avoid providing stimulus information from LED lights and the influence of environmental air. We used a thermographic camera (Avionics/InfReC R450, temperature resolution: $\leq$0.025~$^\circ$C) to capture the skin temperature change. A medium-temperature warm plate (NISSIN/NHP-M20) was used to control the initial skin temperature. In Experiments 2 and 3, the participants wore a pair of noise-canceling headphones and heard white noise to avoid the influence of the compressor, solenoid valve, and environmental noise. We chose to provide stimuli perpendicular to the palm of the right hand (i.e., glabrous, non-hairy skin). As shown in Fig.~\ref{fig:Experiment1ConditionsAndSettings}, we set two hand-fixing bases with holes (circles with a radius of 25 mm) to ensure that the same area of the skin received warm and cold stimuli: base 1 faced the thermographic camera, and base 2 faced the presentation component. To prevent the hand-fixing base from absorbing the heat of light and warming up, we attached aluminum tape (Nitoms/J3230) to the surface facing the presentation component. The environmental temperature was set to 24~$^\circ$C with an air conditioner, whereas the temperature of the generated cold air was 0~$^\circ$C from the measurement using a temperature sensor (Sensirion/SHT85).

\subsection{Experiment 1: Measurement of Temperature Change}
\subsubsection{Experimental procedure}
\begin{figure}[b]
	\centering
 \includegraphics[scale=0.32]{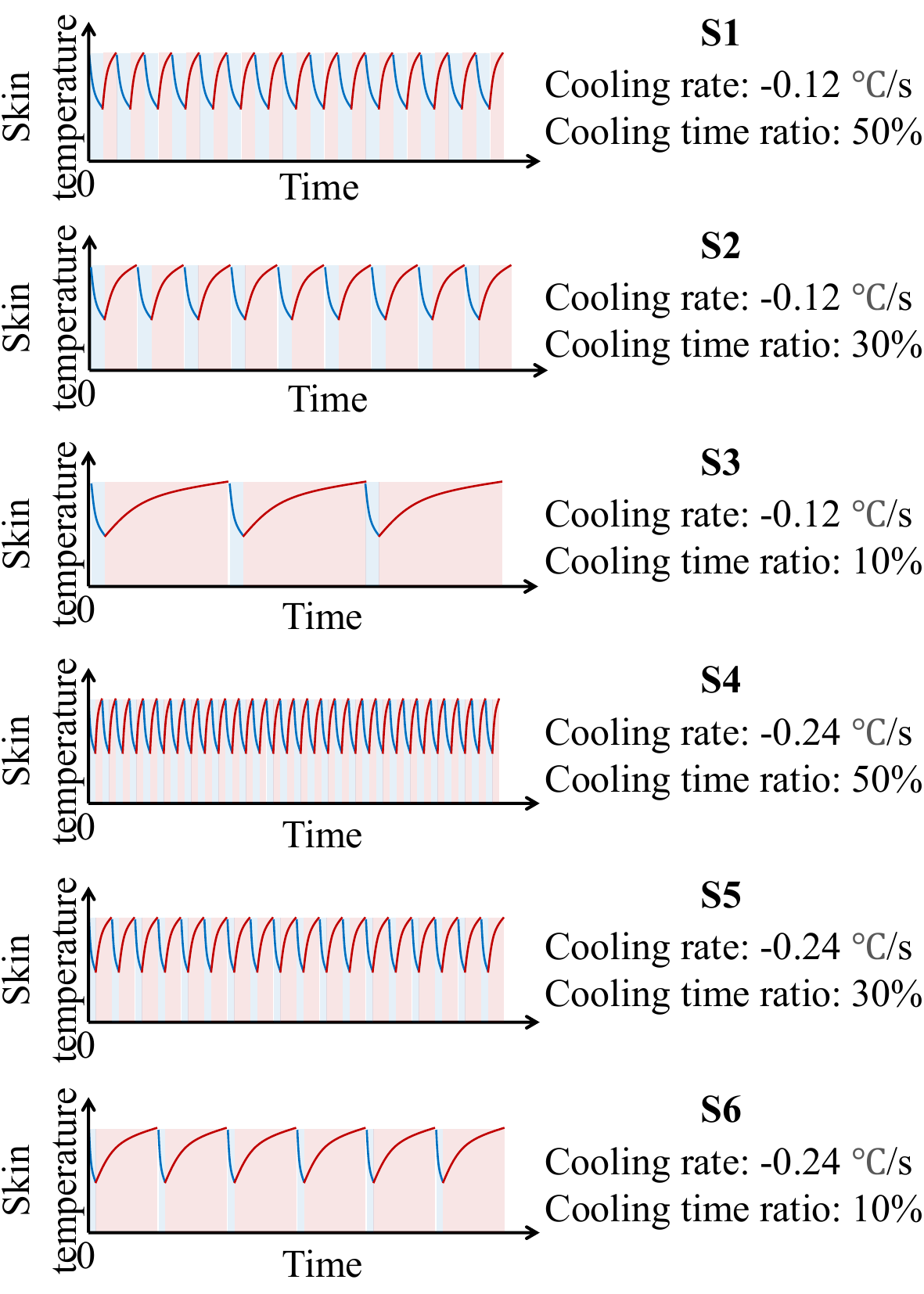}
	\caption{\textcolor{black}{Stimulus pattern for Experiment 1.}}
	\label{fig:designedStimuli1}
\end{figure}

\begin{figure}[b]
	\centering
	\includegraphics[scale=0.32]{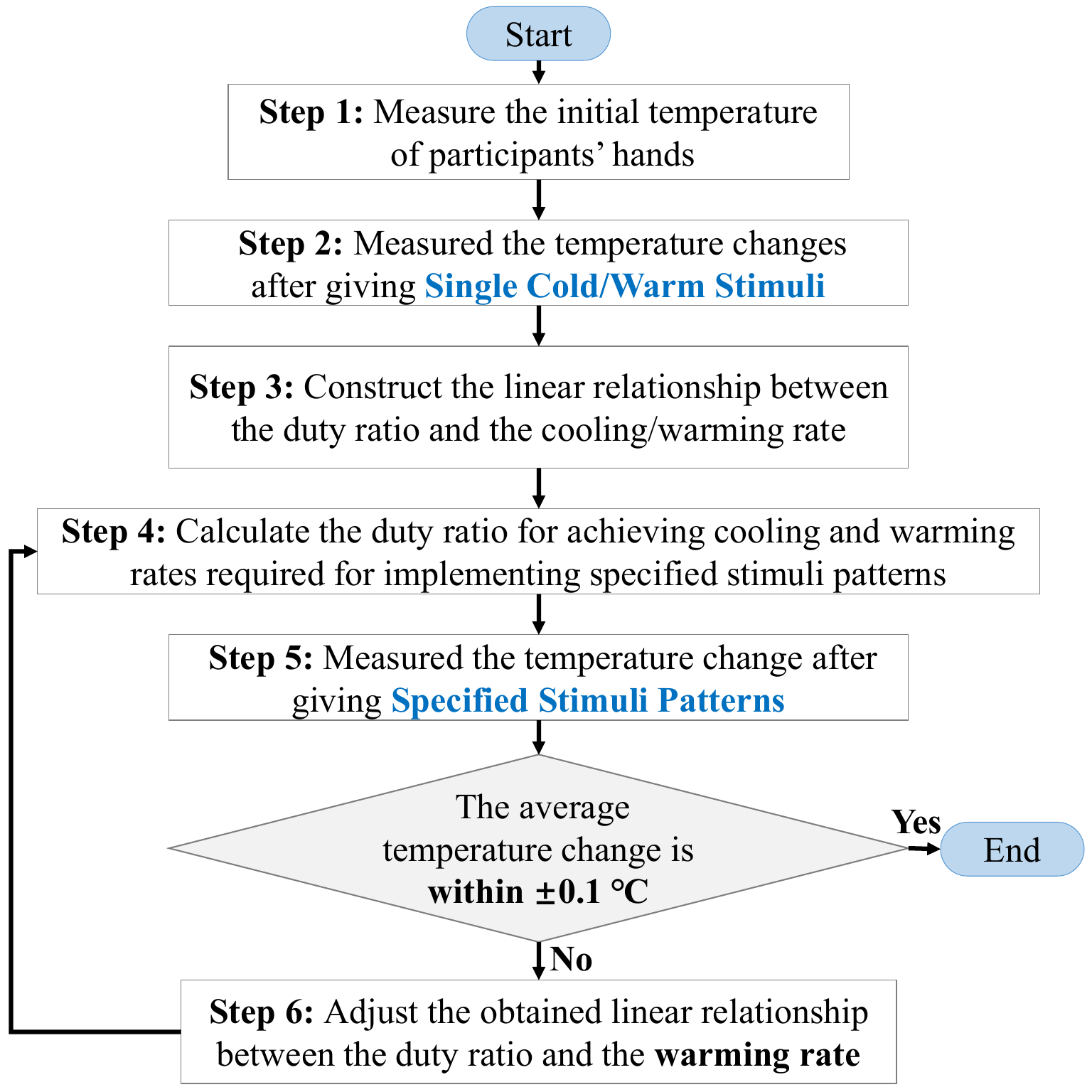}
	\caption{Experimental Flow of Experiment 1 }	\label{fig:flowchartOfEx1}
\end{figure}

\textcolor{black}{
The prototype system lacks real-time monitoring capabilities for tracking changes in skin temperature. Consequently, our initial step involves measuring temperature fluctuations on the skin when applying a single cold or warm stimulus at different output duty ratios. Using these measurements, we establish the relationship between the duty ratio and the cooling/warming rate.
Next, based on the obtained relationship, we control the system output using the stimulus patterns illustrated in Fig.~\ref{fig:designedStimuli1}. We verify whether the skin temperature remains constant before and after the application of these stimulus patterns. In instances where noticeable changes in skin temperature are observed, we proceed with additional calibration to refine the relationship between the duty ratio and warming rate. This calibration process continues until the skin temperature remains nearly unchanged prior to and following the application of these stimulus patterns.
Additionally, considering that individual differences may influence the perception or response to the stimulation, we need to determine the specific PWM duty ratio for each participant. This will ensure that the system's output can implement the intended stimulus patterns for each participant. The specific experimental flow is illustrated in Fig.~\ref{fig:flowchartOfEx1}: }

\begin{enumerate}[Step 1:]
    \item Before the measurement experiments, we requested the participants to raise their right hands without touching anything for 5 min and then measured the initial temperature change of their hands. 
    \item \textcolor{black}{We measured the temperature change $\Delta T_{\rm i}$ after applying a single cold or warm stimulus at a specific duty ratio $D_{\rm i}$ for a duration $\Delta t$ of 6 s. } For cold stimuli, the duty ratios applied to the solenoid valve were set to 49.0, 51.4, 53.8, 56.1, 58.4, and 60.1~\%, whereas the controllable range of the duty ratio was from 49.0 to 60.1~\%. The highest and lowest duty ratios were measured three times, and the other duty ratios were measured once, resulting in ten measurements. For warm stimuli, the duty ratios applied to the LED were set to 11.8, 27.5, 43.1, 58.8, 74.5, and 90.2~\%, while the duty ratio corresponding to the weakest warm stimulus was 11.8~\%, which can cause a rise in skin temperature. The highest warm stimulus that could be tolerated by all participants was below 90.2~\%. The highest and lowest duty ratios were measured three times, and the other duty ratios were measured once, resulting in ten measurements.
    \item \textcolor{black}{We calculated the average of the measured temperature change $\Delta T_{\rm i}/\Delta t$ as the cooling rate $v_{\rm c}$ or warming rate $v_{\rm h}$ at a specific duty ratio $D_{\rm i}$, as follows: 
    \begin{equation}\label{eq:average}
        v_x = T_{\rm i}/\Delta t  \quad (x = \rm{c, h}).
    \end{equation}
    Next, we assumed a linear relationship between the duty ratio $D$ and the cooling/warming rate. The calculated cooling/warming rate using a regression model is expressed as:
    \begin{equation}\label{eq:linear relation}
        \hat{v_x} = a_xD + b_x \quad (x = \rm{c, h}), 
    \end{equation}
    where $a_x$ and $b_x$ are the coefficients.
    To determine the coefficients, 
    we conducted a linear regression analysis. Estimated values $\Tilde{a_x}$ and $\Tilde{b_x}$ can be obtained by
    \begin{equation}
        (\Tilde{a_x},\Tilde{b_x})=\operatorname*{argmin}_{(a_x,b_x)}  \sum_i \left( v_x - \hat{v_x} \right)^2 .
    \end{equation}
    \item Based on the obtained linear relationship (\ref{eq:linear relation}), we set the system output duty ratios to implement the specific stimulus patterns illustrated in Fig.~\ref{fig:designedStimuli1} by achieving the desired cooling or warming rate. The presentation time of each stimulus pattern was 15 s, and the skin temperature change in a cycle was within 0.06~$^\circ$C. }
    \item \textcolor{black}{We measured the temperature changes after giving stimulus patterns. If the average temperature change before and after the given stimulus patterns was within $\pm$ 0.1~$^\circ$C, which is the precision of the analysis program (InfReC Analyzer NS9500 Professional) for the thermal camera, we ended this experiment.}
    \item If the average temperature change before and after the stimulus patterns exceeded $\pm$ 0.1~$^\circ$C, we calibrated the obtained linear relationship between the duty ratio and the warming rate and returned to step~4 to iterate the process again.
\end{enumerate}

\begin{figure}[t]
	\centering
	\includegraphics[scale=0.29]{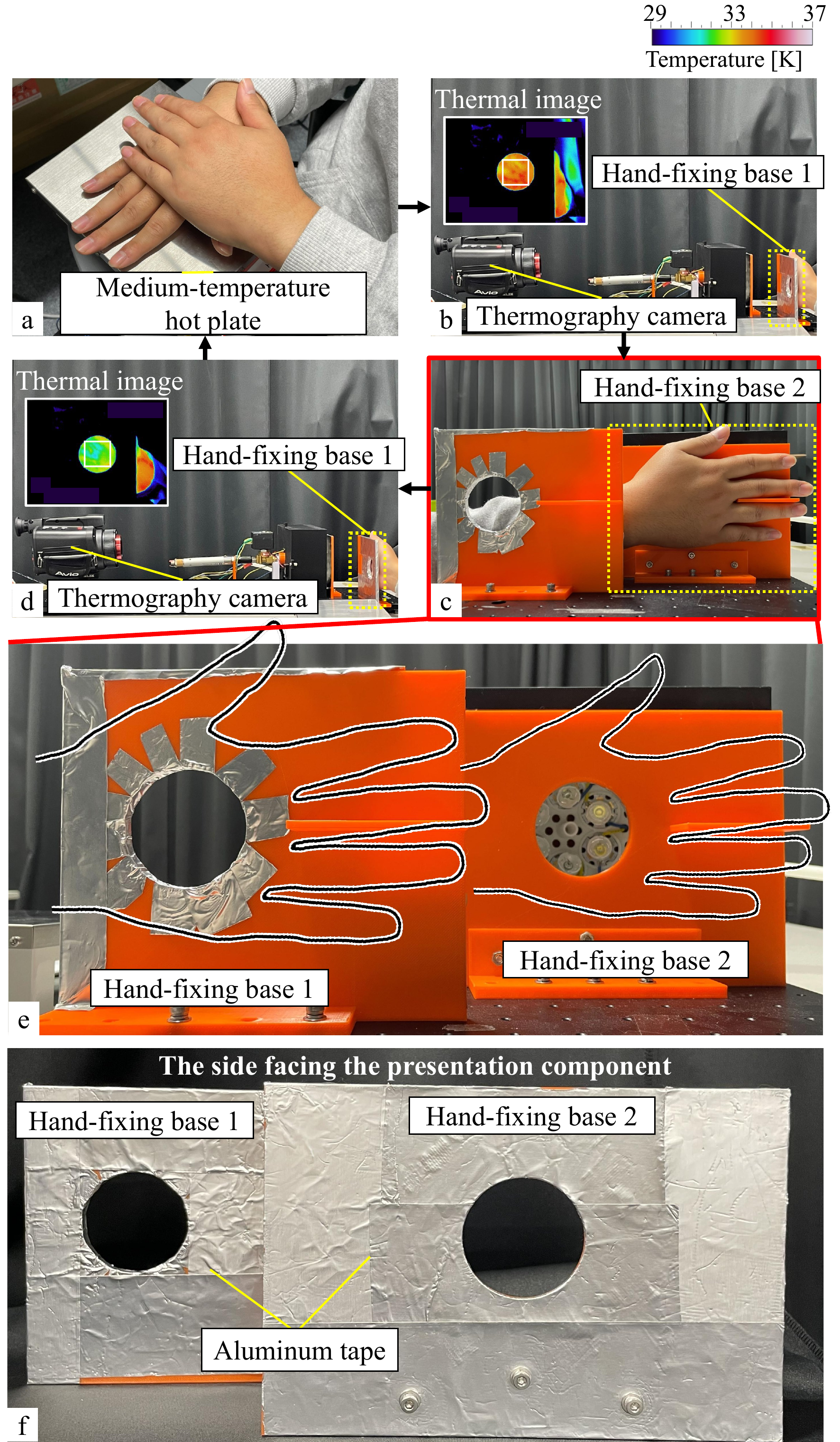}	\caption{Measurement procedure. \textcolor{black}{(a) First, participants placed their hands on a medium-temperature hot plate for approximately 20 s  to initialize the skin temperature. (b) Next, they placed their hands on the hand-fixing base 1 to take a thermal image using a thermographic camera. (c) Following, they put their hands on the hand-fixing base 2 to receive the stimulus. (d) Finally, they put their hands back on the hand-fixing base 1 to retake a thermal image. (e) Positional relationship between the hand-fixing bases 1 and 2. (f) Side facing the presentation component}}	\label{fig:Experiment1ConditionsAndSettings}
\end{figure}
\par The measurement procedure for steps 2 and 5 is illustrated in Fig~\ref{fig:Experiment1ConditionsAndSettings}:
\begin{enumerate}[(a)]
    \item Participants placed their hands on a medium-temperature hot plate, which was set to the initial skin temperature of the participants, for approximately 20 s to initialize the skin temperature. 
    \item Participants placed their hands on the hand-fixing base 1 to take a thermal image using a thermographic camera. We calculated the average skin temperature within the white square (skin area of approximately 35.4 $\times$ 35.4 mm) in the thermal image. 
    \item Participants put their hands on the hand-fixing base 2 to receive the stimulus. \textcolor{black}{At the end of the stimulus, a piezoelectric buzzer (SPL (Hong Kong)/SPT08) was used to play a sound, indicating to participants to immediately move their hands to hand-fixing base 1. This prompt was implemented to minimize potential measurement errors that might arise if participants delayed hand movement after the stimulus concluded.}
    \item Participants put their hands back on the hand-fixing base 1 to retake a thermal image. \textcolor{black}{Furthermore, if the stimulus given in (c) was a single warm stimulus, participants were instructed to hold a bottle of water at room temperature of 24~$^\circ$C for approximately 20 s to cool down before proceeding with the subsequent measurement.}
\end{enumerate}

\subsubsection{Results}
\par
\begin{figure}[htbp]
	\centering
	\subfloat[]
 {\includegraphics[scale=0.32]{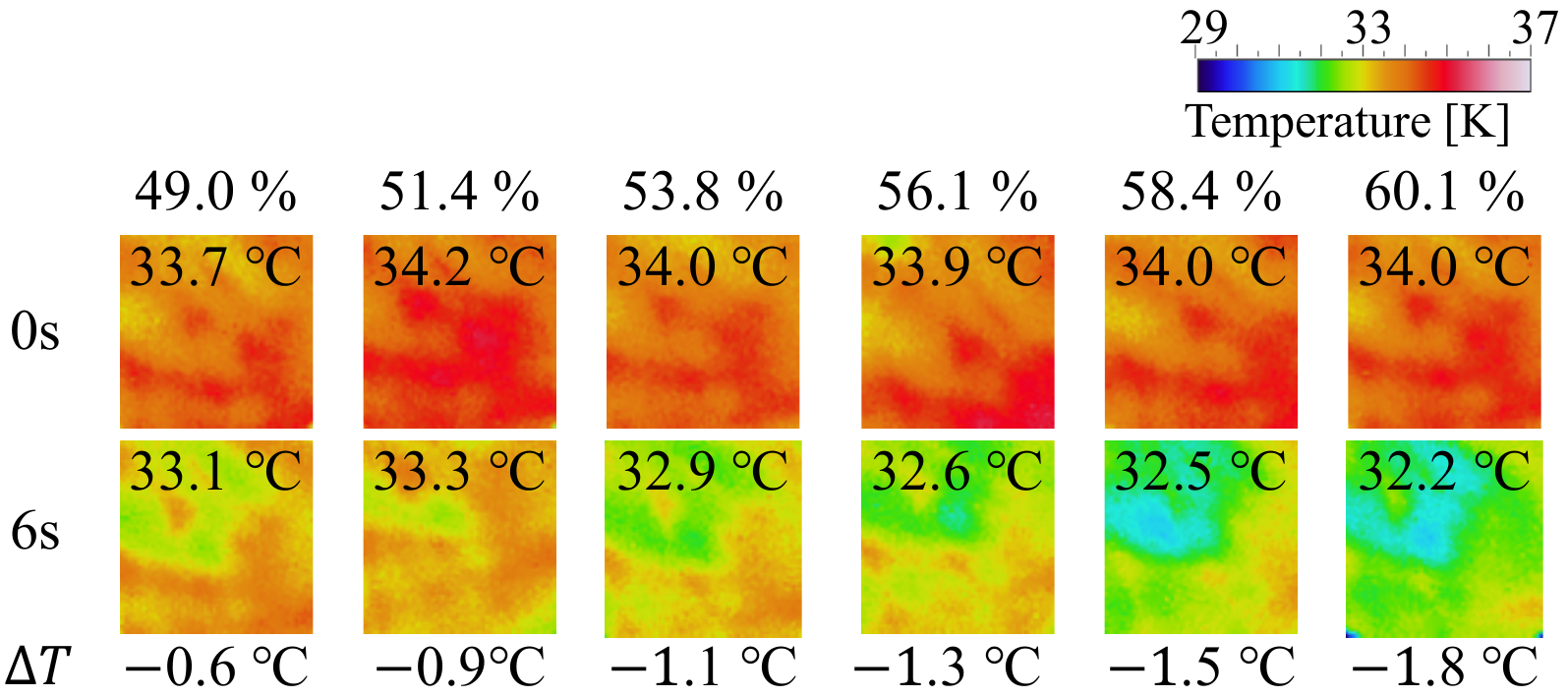}}\\
  \subfloat[]
   {\includegraphics[scale=0.32]{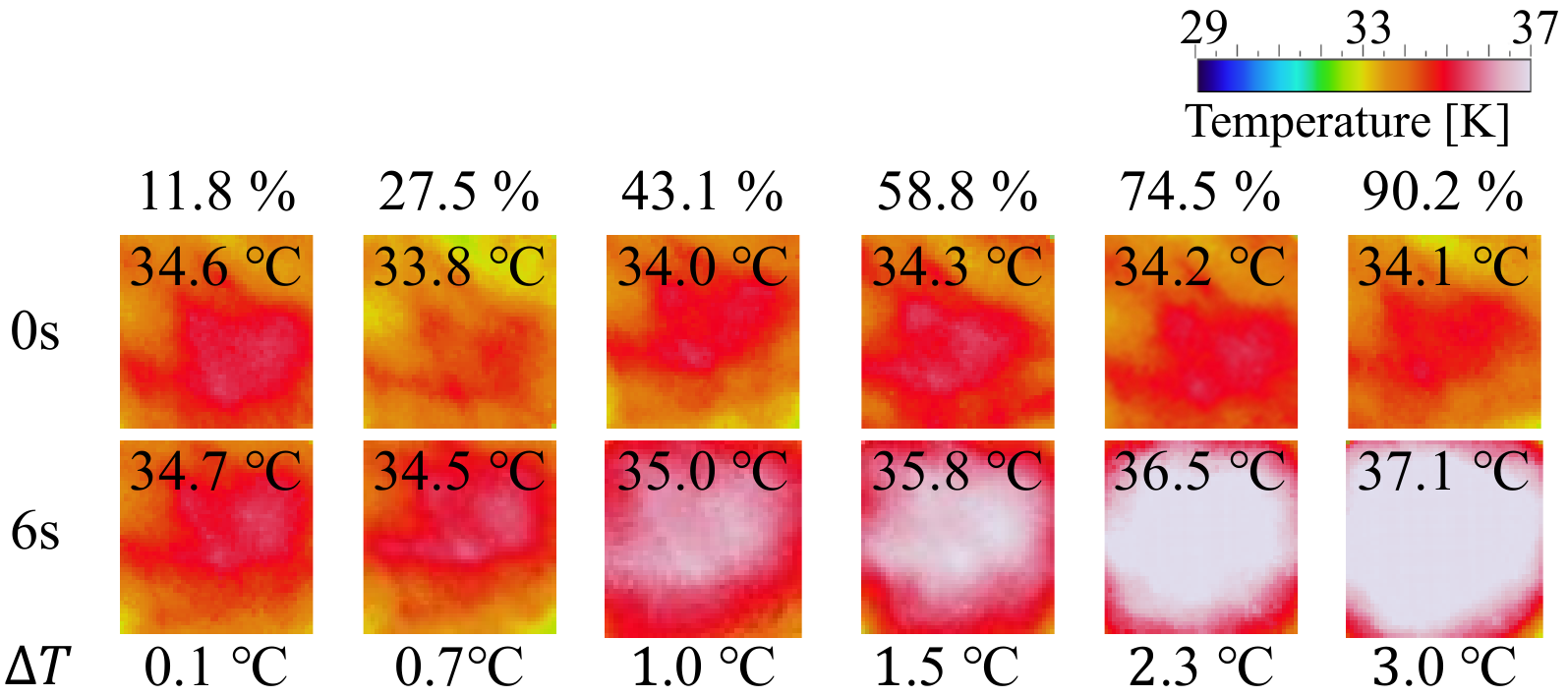}}
	
	\caption{\textcolor{black}{Measurement results in step 2: Example of the thermal images at each duty ratio for (a) cold stimuli. (b) warm stimuli.}}
	\label{fig:step2_result}
	\centering
	\subfloat[]
 {\includegraphics[scale=0.55]{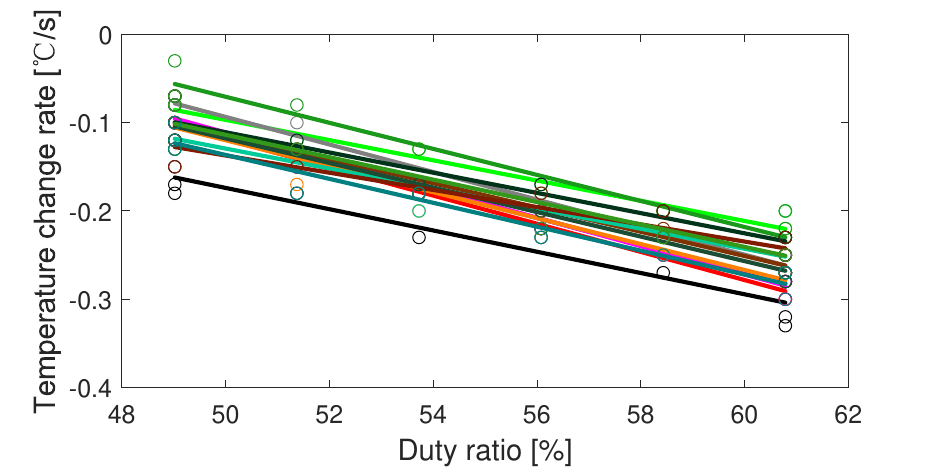}}
 	\\
  \subfloat[]
  {\includegraphics[scale=0.55]{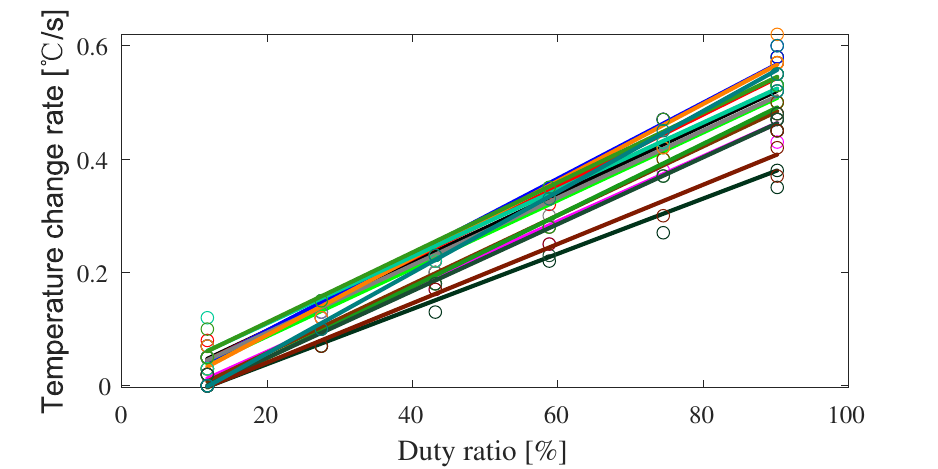}}
	
	\caption{\textcolor{black}{Results of the linear regression analysis conducted in step 3: Linear relationship between (a) the cools rate and the duty ratio. (b) the warming rate and the duty ratio.}}
	\label{fig:step3_result}
 \end{figure}

 Fig.~\ref{fig:step2_result}~(a) and (b) shows an example of the thermal images captured before and after applying a single cold/warm stimulus at each duty ratio in step 2. Fig.~\ref{fig:step3_result}~(a) and (b) present the results of the linear regression analysis conducted in step 3. Each line represents a regression line for an individual participant. The mean coefficient of determination ($R^2$) for the linear regression analysis across all participants was 0.94~$\pm$~0.05 for cold stimuli and 0.98~$\pm$~0.01 for warm stimuli. These results indicate that the cooling/warming rate exhibits a linear relationship with the duty ratio, as we assumed.
\par
\begin{figure}[htbp]
	\centering
 \subfloat[]{\includegraphics[scale=0.34]{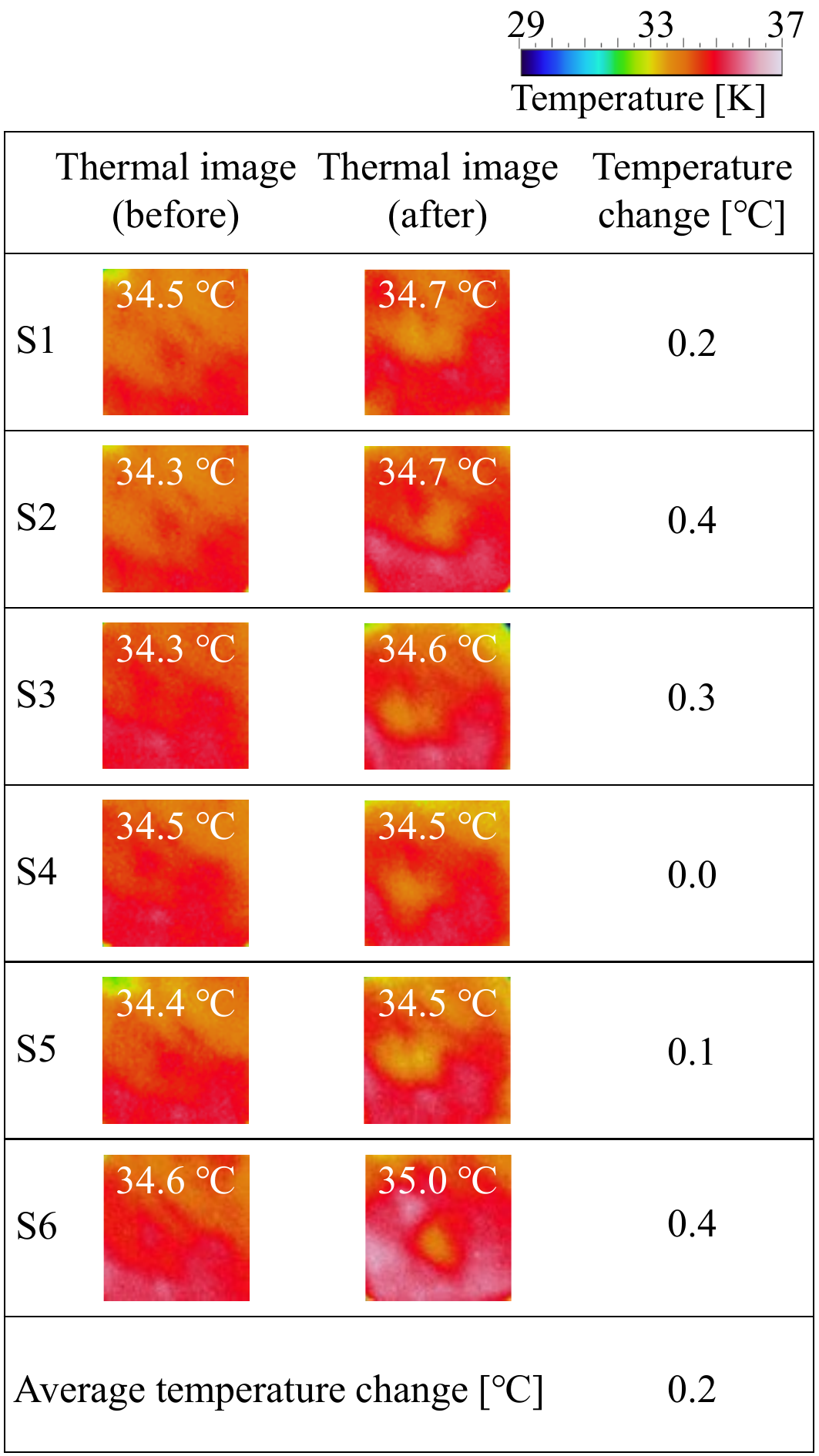}}	\\
 \subfloat[]{\includegraphics[scale=0.34]{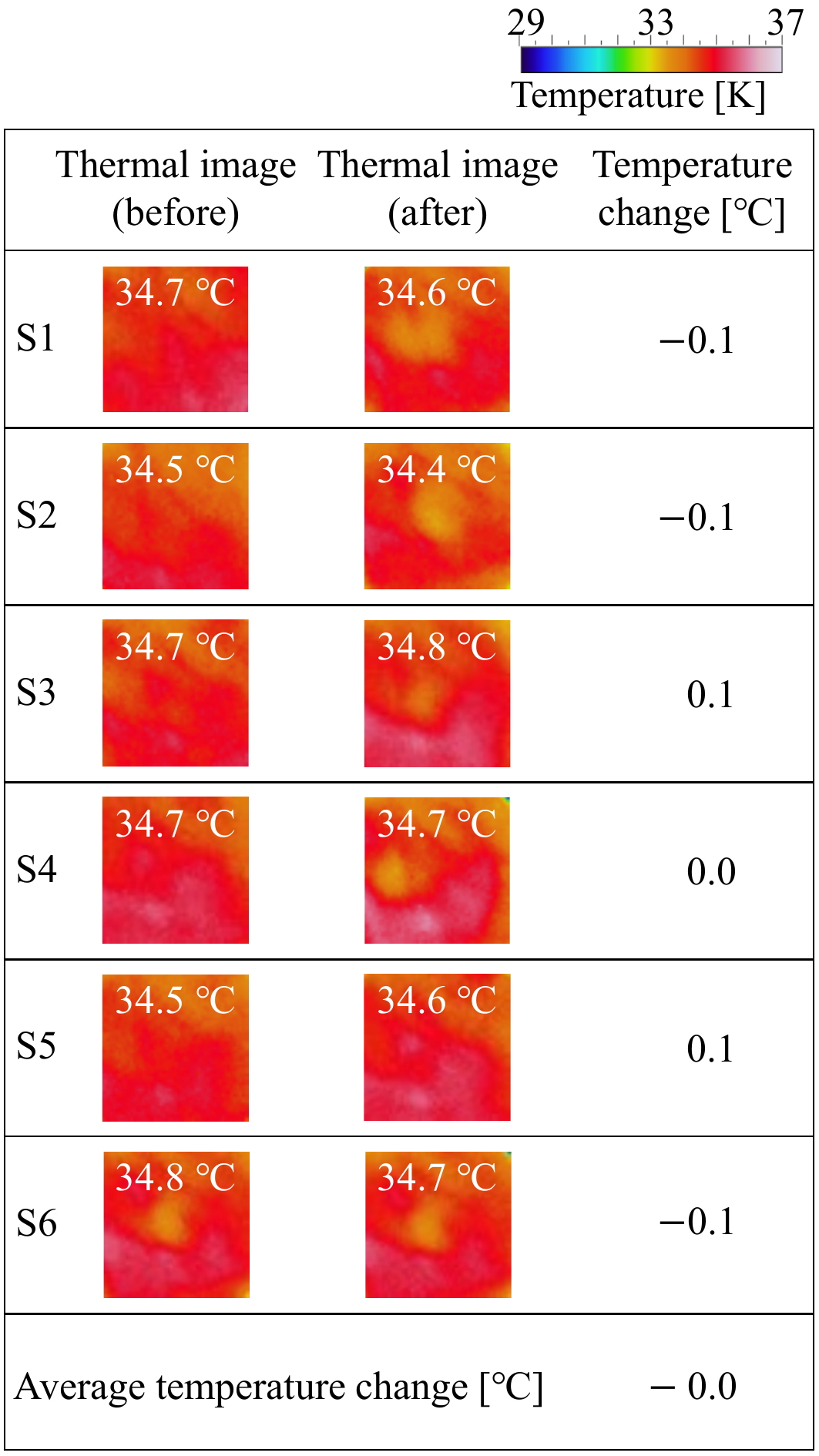}}
	\caption{\textcolor{black}{Measurement results in Step 5: Example of the measurement results (a) before adjustment. (b) after adjustment.}}
	\label{fig:step5_result}
\end{figure}

 Fig.~\ref{fig:step5_result}~(a) shows an example of the measurement results obtained in step 5. These results suggest that the duty ratio assigned to the warm stimulus was excessively high, leading to an undesired rise in the skin temperature. Since the average temperature change for all stimulus patterns exceeded $\pm$~0.1~$^\circ$C, we proceeded to step 6, involve adjusting the obtained linear relationship for the warm stimulus to address the issue of excessive warming. 
Specifically, as the average temperature change was 0.2~$^\circ$C, and the presentation time of each stimulus patterns was given for 15 s, it can be inferred that the current warm stimulus resulted in an additional temperature change rate of 0.2~$^\circ$C/15 s = 0.013~$^\circ$C/s. We added 0.013~$^\circ$C/s to the calculated warming rates in step 3, and performed a new linear regression analysis. Then, based on the updated linear relationship for the warm stimulus, we restarted step 4, repeating the process until the average temperature change for all stimulus types within $\pm$0.1℃. Fig.~\ref{fig:step5_result}~(b) shows the measurement result after adjustment, and the average temperature change is approximately -0.0~$^\circ$C. Since the average temperature was not greater than $\pm$~0.1~$^\circ$C, the adjustment was ended.

\subsection{Experiment 2: Verification of Persistence of Cold Sensation}
\subsubsection{Experimental procedure}
\par
\begin{figure*}[t]
    \centering
\includegraphics[scale=0.3]{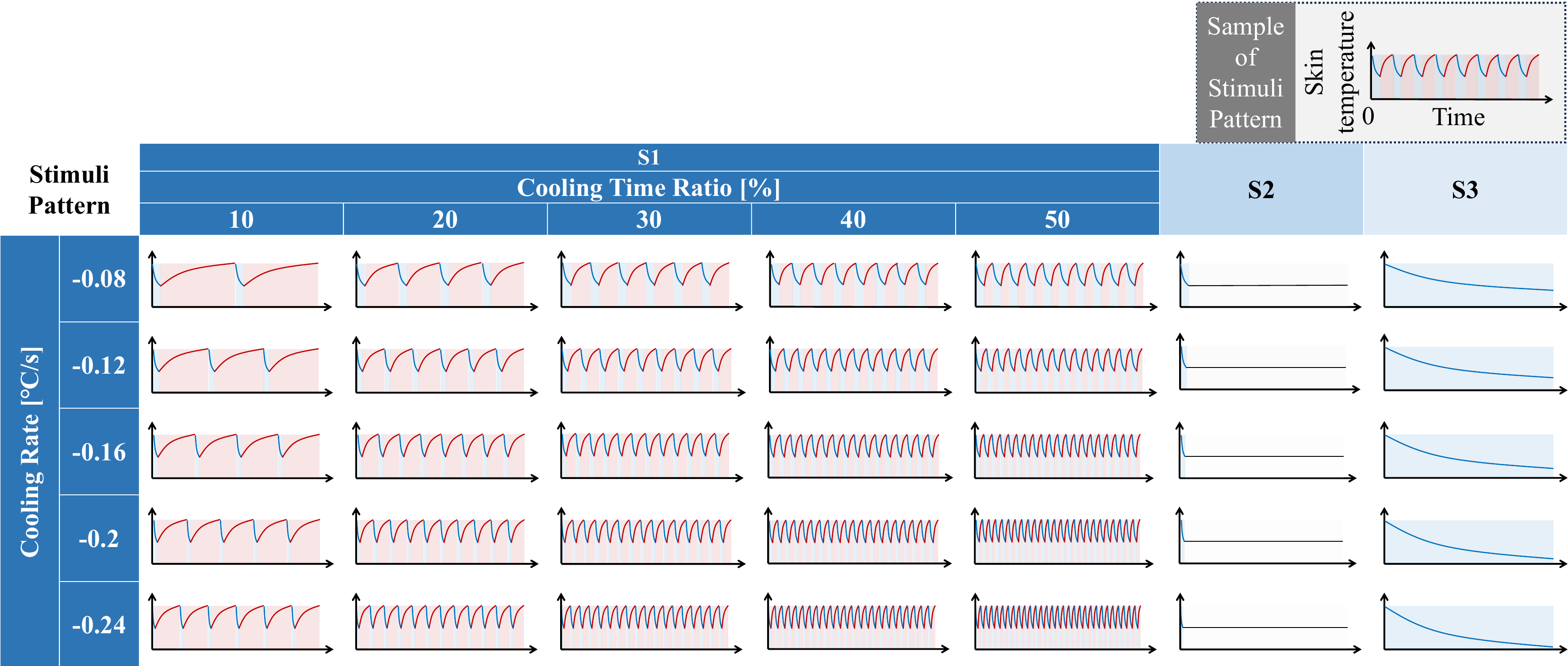}
    \caption{\textcolor{black}{Stimulus pattern for Experiment 2.}}
    \label{fig:ex2}
\end{figure*}

 As shown in Fig.~\ref{fig:ex2}, we varied the skin temperature within 0.06~$^\circ$C and designed 25 stimulus patterns (S1) according to different cooling rates and cooling time ratios. To verify whether the cold sensation elicited by the proposed method was the same as that elicited by the initial temperature drop, we designed stimulus type S2 for comparison. This pattern maintained the decreased temperature by continuously providing warm and cold stimuli of the same intensity after the initial temperature drop was achieved. In addition, for comparison with traditional methods, we designed stimulus type S3. In this pattern, only cold stimuli are applied to the skin, resulting in a continuous decrease in the skin temperature. For each of the stimulus types, S2 and S3, we designed five stimulus patterns based on different cooling rates. The presentation time of each stimulus pattern was 15 s. For each participant, we presented each designed stimulus at random for three trials, resulting in 35 stimuli $\times$ 3 trials = 105 trials. 

 \par
\begin{figure}[htbp]
	\centering
	\includegraphics[scale=0.31]{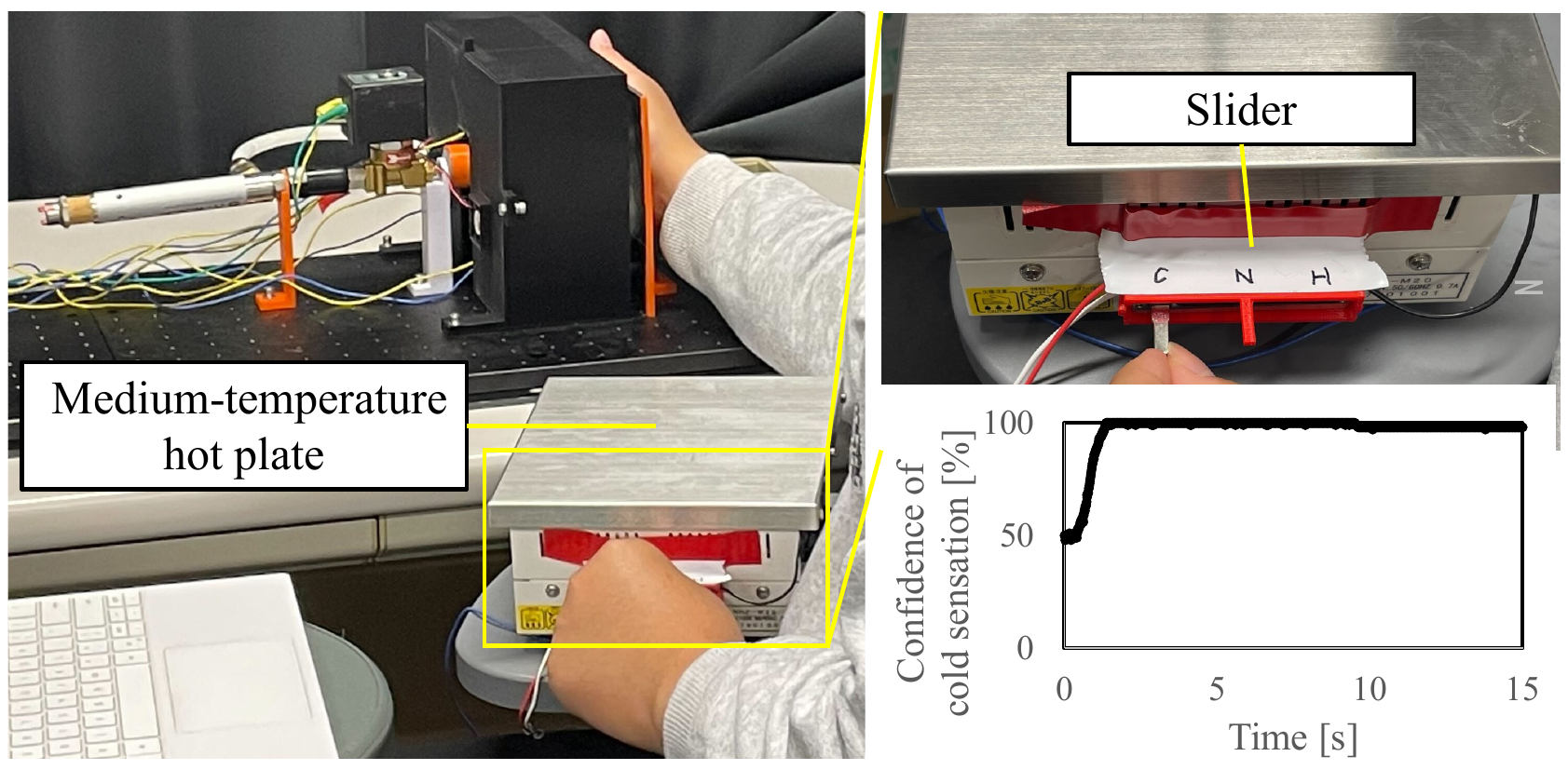}
	\caption{Experimental setting for Experiment 2. Participants responded to the perceived thermal sensation using a slider.}
	\label{fig:ex2Procedure}
\end{figure}

 \textcolor{black}{In each trial, as shown in Fig~\ref{fig:ex2Procedure}, during the stimuli presentation, participants were instructed to make real-time responses based on their sensations using a slider (Supertech Electronic/SL4515N-B10K) positioned on the medium-temperature hot plate. The slider records the participants’ responses at a frequency of 100 Hz. Participants were instructed to adjust the slider when there was a change in the certainty of thermal sensations and focus on perceiving coldness or warmth instead of intensity.  } 
 In cases where participants were confident that they felt cold (warm), move the slider to “C” (“H”); in cases where they were not certain, move the slider between “C” (“H”) and “N” in accordance with their level of confidence. 
 \textcolor{black}{For example, when the confidence level is at 50~\%, move the slider to the midpoint between "C" ("H") and "N".}
 Based on the positions of the sliders, we calculated the participants’ confidence that they felt cold: 100~\% for the "C" end, 50~\% for the "N" end, and 0~\% for the "H" end. For each stimulus, we recorded the dynamic changes in participants' confidence that they felt cold during the 15 s stimulus.
 \par \textcolor{black}{To ensure they understood this instruction, we asked participants to conduct three test trials before the formal experiment, using predefined temperature change patterns (stimulus patterns) to facilitate clear perception of these thermal changes. The followings are the details of the test trials:}
 \begin{itemize}
     \item \textcolor{black}{One of the patterns involved a consistent decrease in skin temperature at a rate of -0.24~$^\circ$C/s over the course of a 15 s stimulus presentation. This pattern caused participants to feel a continuous sensation of cold. We instruct participants to keep the slider at "C" (Cold) side in this situation. }
      \item \textcolor{black}{Another pattern featured a continuous increase in skin temperature at a rate of +0.24~$^\circ$C/s during the same 15 s period. In this case, participants experienced a continuous warm sensation. We instruct participants to keep the slider at the "H" (Warm) side accordingly in this situation.}
      \item \textcolor{black}{The third pattern consisted of fluctuating temperatures. In the first 5 s, the temperature decreased at a rate of -0.24~$^\circ$C/s, then it increased at a rate of +0.24~$^\circ$C/s during the next 5 s interval, followed by another decrease at a rate of -0.24~$^\circ$C/s for the final 5 s. This pattern induced notable variations in skin temperature, facilitating participants in distinguishing transitions between cold and warm sensations. We instructed them to adjust the slider in response to the changes in the certainty of thermal sensations.}
 \end{itemize}
 \begin{figure*}[b]
    \centering
    \subfloat[]
    {\includegraphics[scale=0.3]{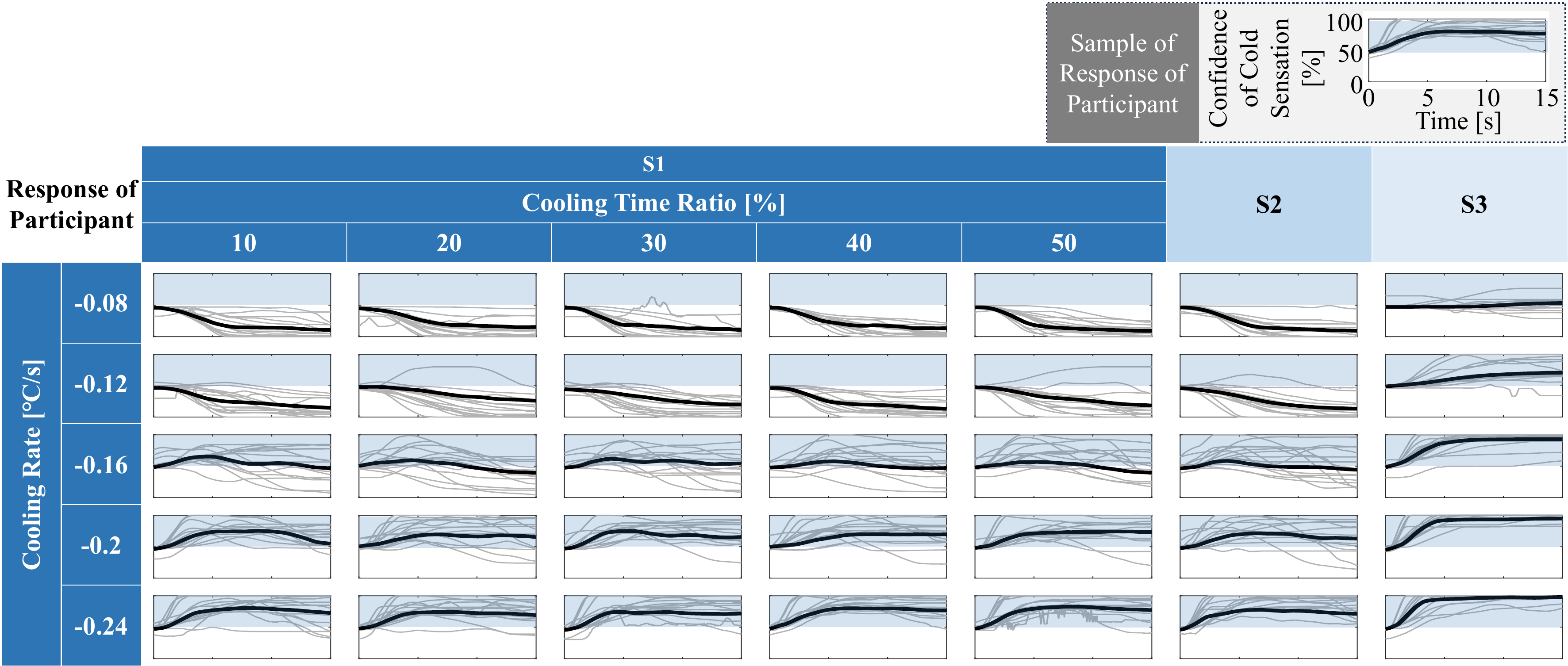}}\\
        \subfloat[]
{\includegraphics[scale=0.3]{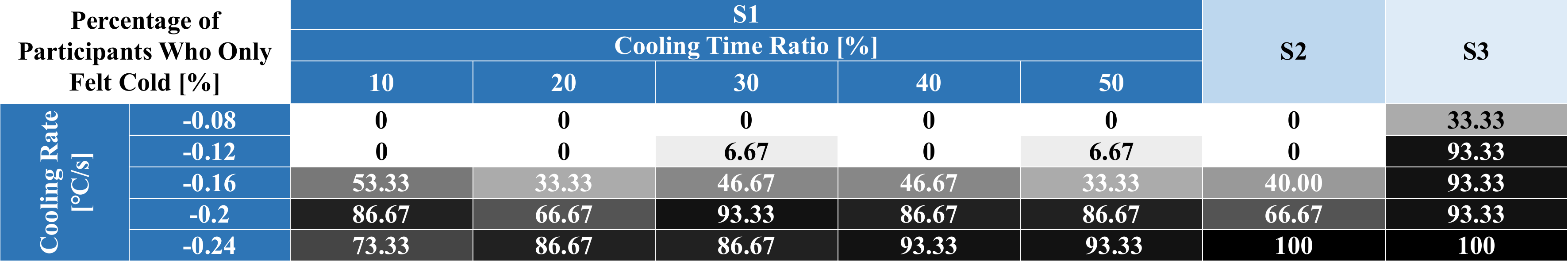}}\\
    \subfloat[]{\includegraphics[scale=0.3]{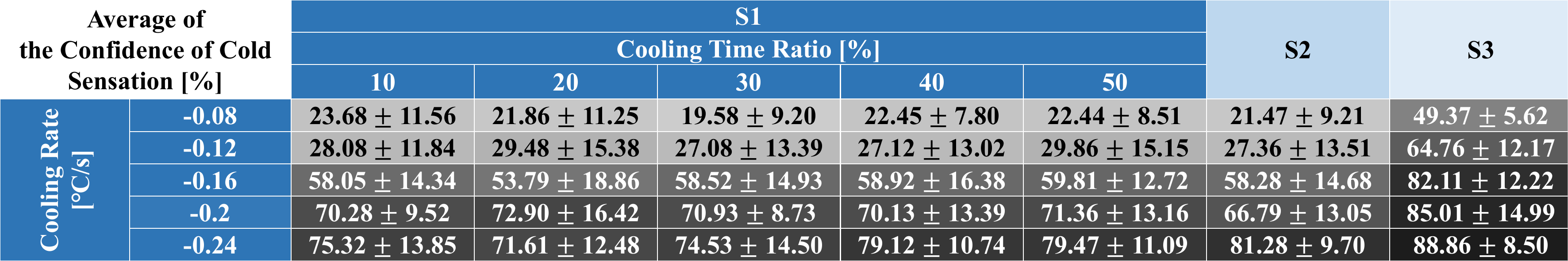}}
    \caption{\textcolor{black}{Results for Experiment 2: (a) Participants responses to each stimulus pattern. (b) Percentage of participants who felt cold persistently without any warmth to each stimulus pattern. (c) Average of the confidence of cold sensation for participants who felt cold persistently to each stimulus pattern.}}
    \label{fig:ex2_result}
\end{figure*}
\subsubsection{Results}
 \par Fig.~\ref{fig:ex2_result}~(a) shows the responses of the participants to each stimulus. Each gray line represents the average response of one participant and the black line represents the average response of all participants. The results show that participants' responses tend to stabilize after 5 s. Accordingly, as shown in Fig. ~\ref{fig:ex2_result}~(b), we calculated the percentage of participants whose confidence levels were all above 50~\% in 5 to 15 s. In these cases, participants felt persistently cold without any warmth. The results indicate that as cooling rates increased, the percentage of participants who felt cold persistently increased as well. When the cooling rate was $-0.24$~$^\circ$C/s, the percentage increased as the cooling time ratio increased. \textcolor{black}{When the cooling rate was within $-$0.2 to $-$0.24~$^\circ$C/s and the cooling time ratio was within 30 to 50~\%, more than 86.67\% of the participants perceived only persistent cold without any warmth. 
 This result demonstrates the effectiveness of our method in inducing cold sensations while maintaining nearly constant skin temperature. This supports our hypothesis that even if there is a period where the skin temperature does not decrease after cold stimuli have been provided, people perceive continuous coldness. Furthermore, the results also indicate that the probability of perceiving continuous cold sensations increases with the alternating rate.} 
 \par \textcolor{black}{As shown in Fig.~\ref{fig:ex2_result}~(c), we calculated the average confidence of cold sensation for each trial in  0–-15 s. In our investigation on stimulus type S1, we analyzed the cooling time ratio and cooling rate effects on the average confidence level of cold perception. According to the non-parametric Kruskal-Wallis test, no statistically significant difference was observed when the cooling time ratio was considered  the influencing factor (\(\chi^2 (4)\) = 0.63, \(p = .960\)). However, when considering the cooling rate as an influencing factor, we found a significant statistical difference (\(\chi^2 (4)\) = 271.34, \(p < .001\)). This suggests that compared to the cooling time ratio, the cooling rate significantly impacts the average confidence level of cold perception for stimulus type S1.}
 \par \textcolor{black}{In subsequent analyses, we expanded our investigation to stimulus types S2 and S3, and likewise carried out the Kruskal-Wallis test with the cooling rate as an influencing factor. For both S2 and S3, we found significant statistical differences. For S2, \(\chi^2 (4)\) = 56.57, \(p < .001\); and for S3, \(\chi^2 (4)\) = 45.06, \(p < .001\). This further verifies that, regardless of the type of stimulus, the cooling rate has a significant effect on the average confidence level of cold perception.
 Next, we combined data from stimulus type S1 with the same cooling rate and conducted the Kruskal-Wallis test with the stimulus type as the influencing factor. For all cooling rates, we observed significant statistical differences. For a cooling rate of -0.08 ~$^\circ$C/s, \(\chi^2 (4)\) = 36.34, \(p < .001\); for -0.12 ~$^\circ$C/s, \(\chi^2 (4)\) = 34.38, \(p < .001\); for -0.16 ~$^\circ$C/s, \(\chi^2 (4)\) = 24.00, \(p < .001\); for -0.20 ~$^\circ$C/s, \(\chi^2 (4)\) = 18.76, \(p < .001\); and for -0.24 ~$^\circ$C/s, \(\chi^2 (4)\) = 14.08, \(p < .001\).}
 \par
\begin{figure}[t]
        \vspace{3mm}
        \centering
        \includegraphics[scale=0.35]{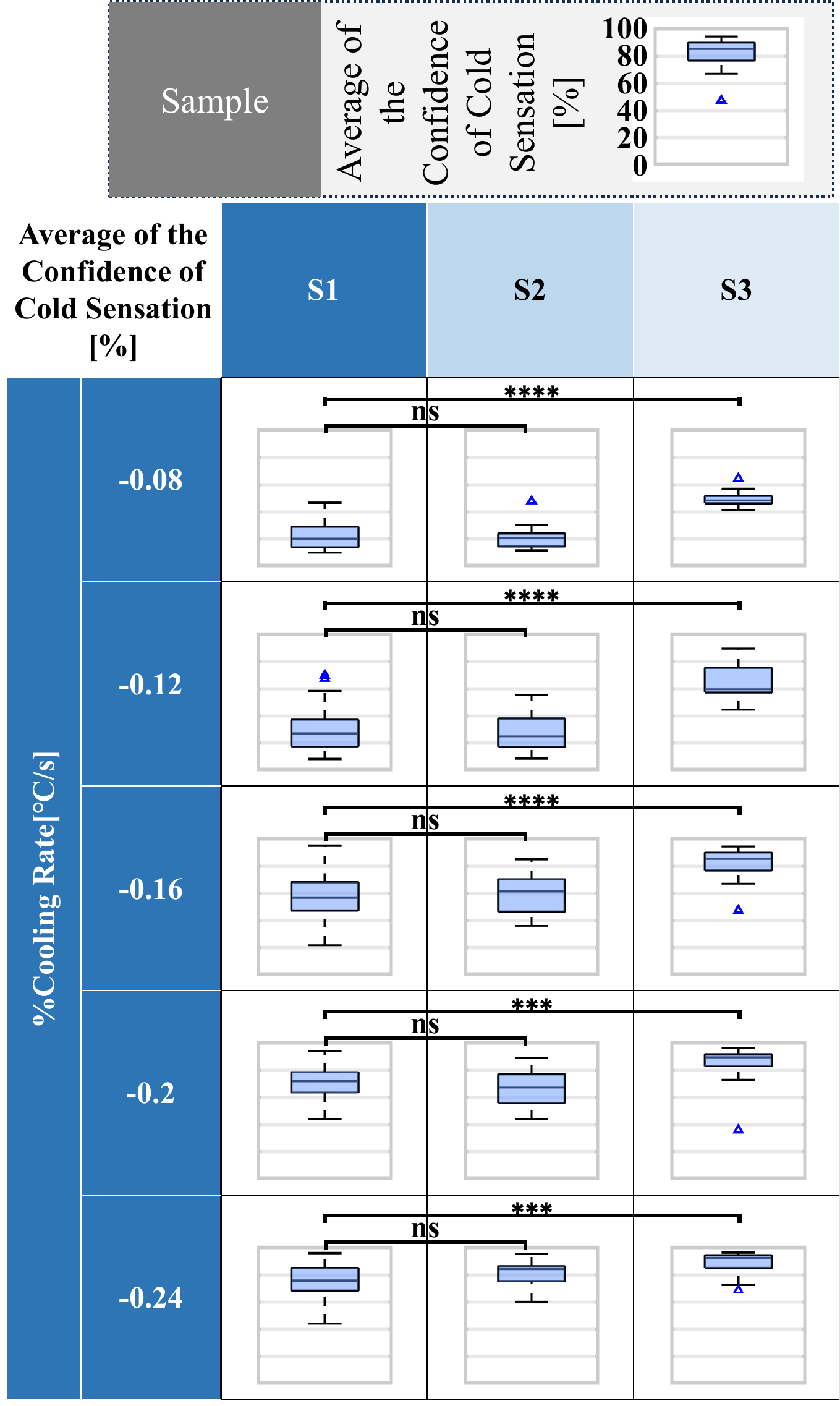}
        \caption{\textcolor{black}{Level of coldness rating to each stimulus pattern (ns:~p~$>$~0.05, ***: p~$<$~0.001, ****: p~$<$~0.0001).}}
        \label{fig:ex2_result_multicompare}  
    \end{figure}

 \textcolor{black}{To further investigate the specific group differences, we conducted pairwise comparisons using the Wilcoxon rank-sum test, and the Benjamini-Hochberg procedure was employed for False Discovery Rate (FDR) correction. In our analysis, we set the FDR at 5~\%, which helps minimize the likelihood of incorrectly identifying significant group differences due to chance. As illustrated in Fig.~\ref{fig:ex2_result_multicompare}, we obtained adjusted p-values for pairwise comparisons by applying the Benjamini-Hochberg method. These adjusted p-values allow us to determine which specific pairs of stimulus types exhibit statistically significant differences in the average confidence level of cold perception while accounting for the FDR. According to the adjusted p-values obtained from the analysis, we found statistically significant differences between S1 and S3 across all cooling rates (adjusted \(p < .001\)). This indicates that at the same cooling rate, the cold perception confidence of stimulus type S1 is lower than that of stimulus type S3, where the temperature continues to drop. On the other hand, we did not find any significant differences between S1 and S2 regardless of the cooling rate: for a cooling rate of -0.08 ℃/s, adjusted \(p > .999\); for -0.12 ℃/s, adjusted \(p = .871\); for -0.16 ℃/s, adjusted \(p = .940\); for -0.20 ℃/s, adjusted \(p = .304\); and for -0.24 ℃/s, adjusted \(p = .116\). The temperature of  stimulus type S2 drops initially, while stimulus S1 maintains the initial skin temperature.}
\subsection{Experiment 3: Investigation of Intensity of Cold Sensation}
\subsubsection{Experimental procedure}
\par
\begin{figure}[t]
        \vspace{3mm}
        \centering
\includegraphics[scale=0.28]{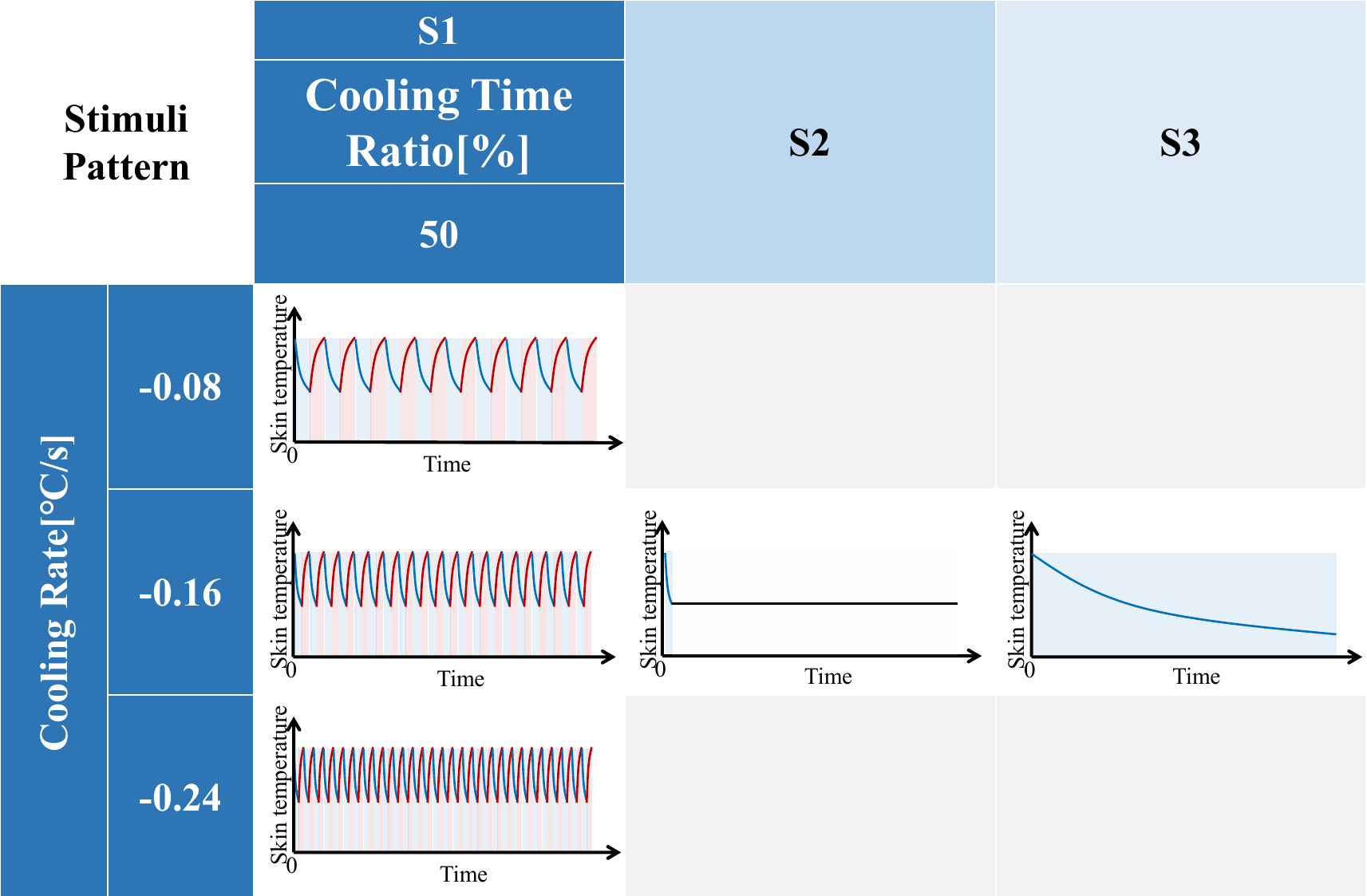}
        \caption{\textcolor{black}{Stimulus pattern for Experiment 3.}}
        \label{fig:ex3}  
    \end{figure}

 \textcolor{black}{The main goal of this experiment was to explore potential variations in cold sensation intensity induced by stimulus type S1, which maintained a relatively constant skin temperature. To facilitate a meaningful comparison, stimulus type S2, involving only an initial temperature decrease, and stimulus type S3, resulting in a continuous decrease in skin temperature, were also included.
The S1 stimulus pattern with a cooling rate of $-$0.16~$^\circ$C/s and a cooling time ratio of 50~\% was selected as the baseline stimulus. Considering the close relationship between cold sensation intensity and the cooling rate~\cite{temperatureSensitivity}, we chose additional S1 stimulus patterns with cooling rates of $-$0.16 and $-$0.24~$^\circ$C/s while keeping the cooling time ratio constant for comparative analysis. Moreover, as shown in Fig.~\ref{fig:ex3}, the S2 and S3 stimulus patterns, both with a cooling rate of $-$0.16~$^\circ$C/s, were included for further comparison.}
The presentation time of each stimulus pattern was 15 s. For each participant, we presented each designed stimulus at random for three trials, resulting in 5~stimuli~$\times$~3~trials~=~15 trials. \textcolor{black}{After each stimulus presentation, a 7-point Likert scale (1 = Not at all cold, 4 = moderately cold, 7 = very cold) was displayed on a PC monitor. We instructed the participants to rate their perceived level of coldness by selecting the appropriate point on the Likert scale using the mouse.}
\subsubsection{Results}
\par
\begin{figure}[t]
        \vspace{3mm}
        \centering
        \includegraphics[scale=0.4]{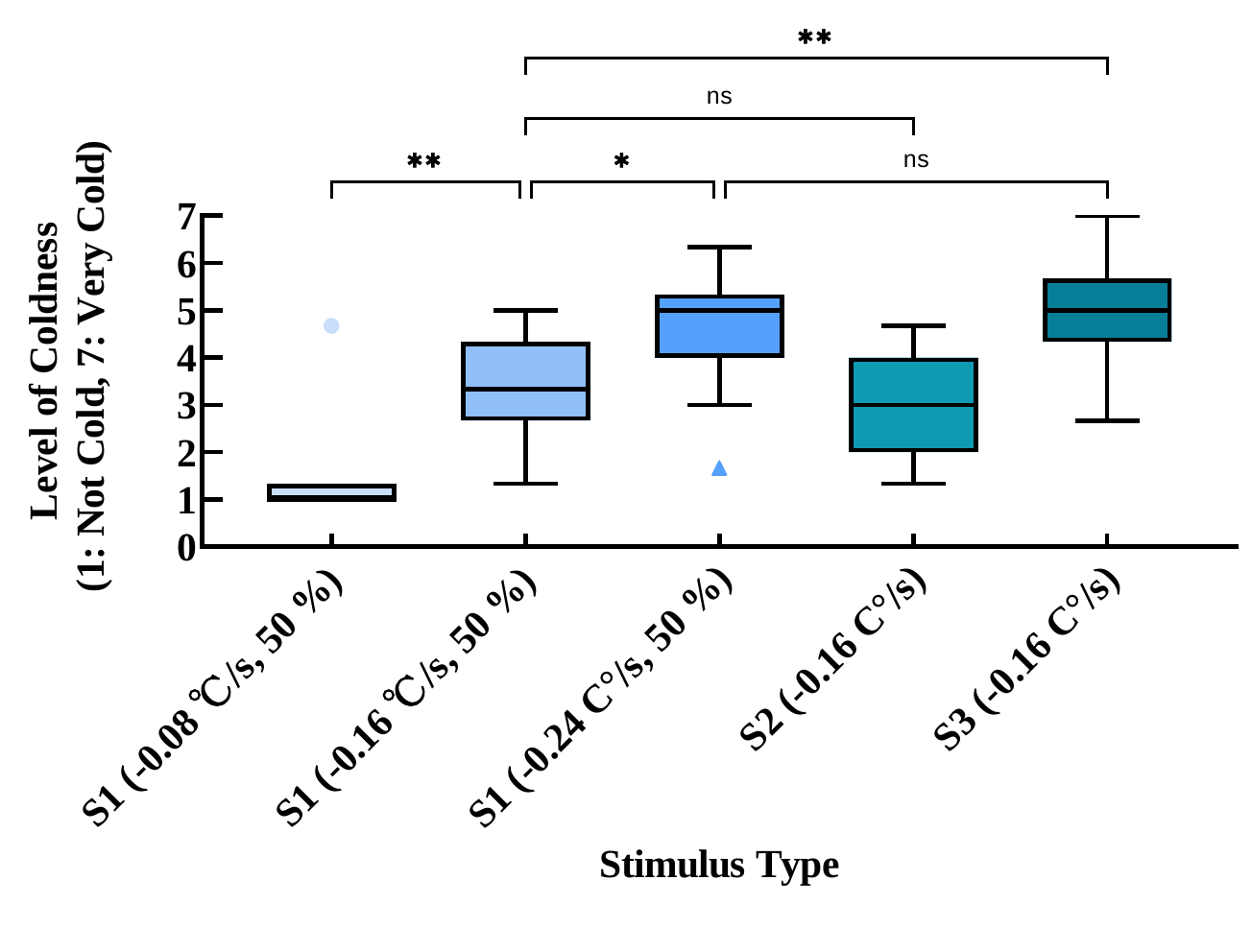}
        \caption{\textcolor{black}{Level of coldness rating to each stimulus pattern (ns:~p~$>$~0.05, *:~p~$<$~0.05, **: p~$<$~0.01).}}
        \label{fig:ex3_result}  
    \end{figure}

 \textcolor{black}{Results were analyzed using the Kruskal-Wallis test. The test showed a significant effect of stimuli type on the level of coldness rating (\(\chi^2 (4)\) = 45.72, \(p < .001\)). To further investigate the specific group differences, we conducted pairwise comparisons using the Wilcoxon rank-sum test, and the Benjamini-Hochberg procedure was employed for False Discovery Rate (FDR) correction. In our analysis, we set the FDR at 5~\%, which helps minimize the likelihood of incorrectly identifying significant group differences due to chance.
As shown in Fig.~\ref{fig:ex3_result}, we obtained adjusted p-values for the pairwise comparisons by applying the Benjamini-Hochberg method. These adjusted p-values allow us to determine which specific pairs of stimulus types exhibit statistically significant differences in coldness rating while accounting for the FDR.
Based on the adjusted p-values obtained from the analysis, we found that the comparison between S1($-0.16$~$^\circ$C/s, 50~\%) and S1($-0.08$~$^\circ$C/s, 50~\%) was statistically significant (adjusted \(p < .001\)). This indicates that participants rated S1($-0.16$~$^\circ$C/s, 50~\%) as significantly colder than S1($-0.08$~$^\circ$C/s, 50~\%). Similarly, we observed a significant difference between S1($-0.16$~$^\circ$C/s, 50~\%) and S1($-0.24$~$^\circ$C/s, 50~\%) (adjusted  \(p = .004\)). Participants rated S1($-0.24$~$^\circ$C/s, 50~\%) as significantly colder compared to S1($-0.16$~$^\circ$C/s, 50~\%). These results indicate that when the cooling time ratio remained unchanged, the cold sensation tended to become stronger with increasing cooling rates. In comparison to stimulus type S2, no significant difference was found between S1($-0.16$~$^\circ$C/s, 50~\%) and S2($-0.16$~$^\circ$C/s) (adjusted  \(p = .464\)). 
In comparison to stimulus type S3, a significant difference was observed between S1 ($-0.16$~$^\circ$C/s, 50~\%) and S3 ($-0.16$~$^\circ$C/s) (adjusted  \(p < .001\)). However, no significant difference was found between S1 ($-0.24$~$^\circ$C/s, 50~\%) and S3 ($-0.16$~$^\circ$C/s) (adjusted  \(p = .372\)).}
 
\section{Discussion}
\par In this study, we present a non-contact persistent cold sensation presentation system that integrates different heat transfer mechanisms to control cold and warm stimuli independently. In particular, we used a cold air source, which relies on convection heat transfer to cool, and a light source, which relies on radiation heat transfer to heat. Using non-contact stimuli, we alternately changed the temperature rise and fall in the same skin area by providing cold stimuli continuously, while intermittently providing warm stimuli. 
\par In Experiment 1, in preparation for Experiments 2 and 3, we first measured the temperature changes due to a single cold or warm stimulus. \textcolor{black}{Using these measurements, we determined the linear relationship between the duty ratio and the cooling/warming rate. Subsequently, we applied stimulus patterns, designed based on this derived linear relationship.
Considering that the implementation of individual temperature assessments for each stimulus pattern in Experiments 2 and 3 would be significantly time-consuming and potentially impose additional burdens on the participants, we abstained from conducting temperature measurements during the execution of these experiments. Instead, we opted to test a subset of stimuli patterns used in Experiments 2 and 3 during Experiment 1, with the objective to validate whether the skin temperature could remain almost constant before and after applying the stimulus patterns. If noticeable changes in skin temperature are observed, we proceeded with additional calibration to refine the relationship between the duty ratio and warming rate, successfully calibrating each participant.} 
\par In Experiment 2, we examined the relationship between the persistence of cold sensation and the cooling rate as well as the cooling time ratio. 
\textcolor{black}{Our findings show that with a cooling rate within $-$0.2 to $-$0.24~$^\circ$C/s and a cooling time ratio within 30 to 50~\%, over 86.67~\% of participants experienced a persistent sense of coldness. This highlights the efficacy of our approach in eliciting cold sensations while maintaining an almost steady skin temperature. These findings affirm our hypothesis that even if there is a period where the skin temperature does not decrease after cold stimuli have been provided, people perceive continuous cold. Moreover, the results signify that the likelihood of perceiving continuous cold sensations escalates in correlation with the alternating rate.} 
In comparison with stimulus type S2, which maintained the initial decreased temperature, stimulus type S1, which maintained the skin temperature almost unchanged, showed almost the same confidence in cold sensations. However, compared with stimulus type S3, which decreased temperature continuously, stimulus type S1 had less confidence in cold sensations. This is because the cold sensation is not solely related to the rate of temperature change but also the amount of change in skin temperature. Conversely, when the cooling rate was within $-$0.08 to $-$0.16~$^\circ$C/s, most participants felt only warm. 
\textcolor{black}{One possible explanation for this phenomenon could be that the intensity of the cold stimuli precisely balances out the natural heat generated by the body, resulting in the absence of cold sensations. In contrast, the warm stimuli contributed to the perception of warmth. Taking advantage of this finding, our method might be able to induce warm sensations while maintaining a constant skin temperature. This will be studied in greater detail in the future.}
\par In Experiment 3, we investigated the intensity of the cold sensation elicited by different stimulus patterns. We found that when the cooling time ratio remained unchanged, the cold sensation tended to become stronger with an increasing cooling rate. In comparison with stimulus type S2, when the cooling rates were the same, the intensities of the cold sensations elicited by stimulus type S1 did not differ significantly. Compared with stimulus type S3, when the cooling rate remained unchanged, the intensity of the cold sensation elicited by the proposed method was weaker. However, the results also indicated that the proposed method could achieve a similar intensity to the traditional method by increasing the cooling rate while maintaining an almost constant skin temperature.
\par In all the experiments, we conducted experiments on the palm. In the future, we will conduct experiments on other parts of the skin that are not covered by clothing, such as the face and neck. In addition, cold air causes pressure sensations on the skin and may affect participants' ability to distinguish thermal sensations. Our future work will also examine the impact of pressure sensations on temperature perception by transforming cold airflow into turbulence using nets. 
\par \textcolor{black}{The prototype system lacks a real-time tracking feature to monitor changes in skin temperature. Due to the hand movements during the measurement process in Experiment 1, there may be slight errors in the recorded skin temperature variations. In the future,} we plan to find a small, high-precision, non-contact temperature sensor. We will develop a non-contact persistent cold sensation presentation system that requires no prior calibration and can monitor and control the changes in the skin temperature in real time by placing this sensor within the display part. 
\textcolor{black}{Additionally, the presentation distance is only 42 mm, and the experimental setup was fixed to a table. In the future, we plan to develop a system that can move the display part along with the user's movements, such as using a mechanical arm, so that users can move freely in a virtual environment. We also consider integrating the display part into existing interactive devices, such as gaming chairs or head-mounted displays (HMDs), for practical use. In the future, we also envision promoting remote thermal sensation by optimizing the shape of the cold air outlet nozzle or the lenses for light focusing, extending the distance that cold air or visible light can travel.
Also, we are considering integrating our method into VR environments, allowing for the simulation of scenarios such as a beach with a gentle breeze or a snowfield with a strong wind. Combining thermal sensations with visual, auditory, and olfactory stimuli, we aim to explore how users perceive thermal stimuli within a multisensory environment.}
\par \textcolor{black}{The contributions of this study are as follows:
\begin{itemize}
    \item We developed a novel thermal display that integrates two independent heat transfer mechanisms and proposed a method to present persistent cold sensations with lower residual heat in a non-contact manner.
    \item We proved that even if there is a period of time where the skin temperature does not decrease after cold stimuli has been provided, persistent cold sensations can be elicited by quickly alternating temperature rise and fall in the same area of the skin.
    \item We investigated the effect of both cooling rate and cooling time ratio on the intensity and persistence of cold sensations.
\end{itemize}
}

\section{Conclusion}
\par In this study, we proposed a method to elicit non-contact cold sensations with low residual heat by integrating two independent heat transfer mechanisms; convection heat transfer for cooling and radiation heat transfer for warming. We found that our method could elicit cold sensations while maintaining a constant skin temperature. When the cooling rate was within $-$0.2 to $-$0.24~$^\circ$C/s and the cooling time ratio was within 30 to 50~\%, more than 86.67~\% of participants felt persistent cold. We also examined the intensity of the cold sensation elicited by the proposed method. According to our results, the intensity of the proposed method did not differ significantly from that of the traditional method when the cooling rate was increased. There is a limitation in that the prototype system requires prior calibration. We plan to find a small, high-precision temperature sensor and develop a system that does not require prior calibration. Additionally, we plan to incorporate our novel method into a multisensory interactive system to provide users with a more immersive experience in applications, such as virtual reality or metaverse.

\begin{IEEEbiography}[{\includegraphics[width=1in,height=1.25in,clip,keepaspectratio]{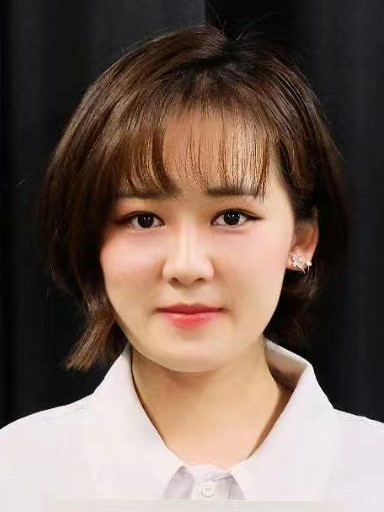}}]{Jiayi Xu} \textcolor{black}{(Member, IEEE) received her Ph.D. degree in engineering from University of Tsukuba, Tsukuba, Japan, in 2023. Since 2023, she has been a Postdoctoral Researcher with the Institute of Systems and Information Engineering at the University of Tsukuba. }Her research interests include haptics with a focus on thermal sensations.

\end{IEEEbiography}

\begin{IEEEbiography}[{\includegraphics[width=1in,height=1.25in,clip,keepaspectratio]{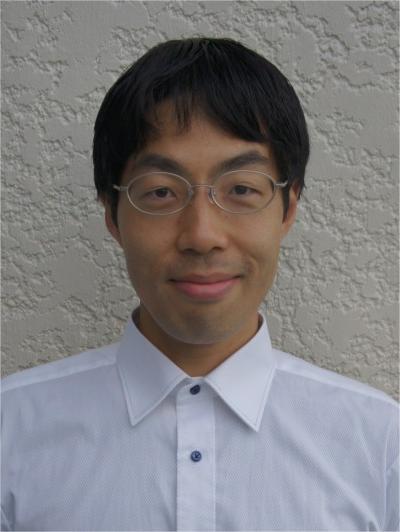}}]{Shoichi Hasegawa} (Member, IEEE) received the D.Eng. degree in computational intelligence and systems from the Tokyo Institute of Technology, Tokyo, Japan. 
He has been an Associate Professor with the Tokyo Institute of Technology since 2010 and was previously an Associate Professor with the University of Electro--Communications. His domain of research includes haptic renderings, realtime simulations, interactive characters, soft and entertainment robotics, and virtual reality.
\end{IEEEbiography}

\begin{IEEEbiography}[{\includegraphics[width=1in,height=1.25in,clip,keepaspectratio]{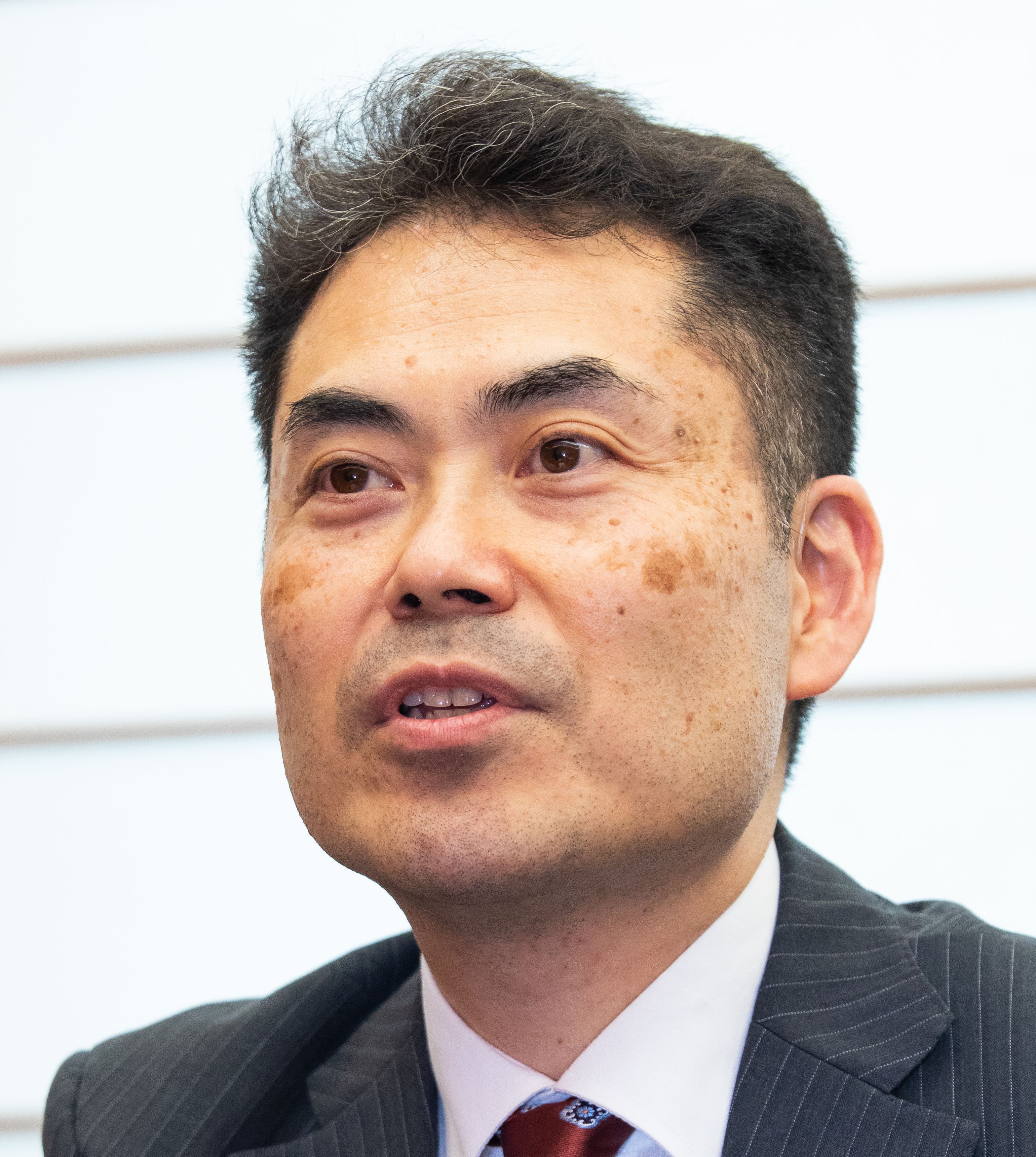}}]{Kiyoshi Kiyokawa} (Member, IEEE) received his Ph.D. degree in information systems from Nara Institute of Science and Technology (NAIST) in 1998. After working at the National Institute of Information and Communications Technology (NICT) and Osaka University, he is currently a Professor at NAIST, leading the Cybernetics and Reality Engineering Laboratory. His research interests include virtual reality, augmented reality, human augmentation, 3D user interfaces, CSCW, and context awareness. He is also a Fellow of the Virtual Reality Society of Japan. He is an associate editor-in-chief of IEEE TVCG and an inductee of the IEEE VGTC Virtual Reality Academy (Inaugural Class).
\end{IEEEbiography}

\begin{IEEEbiography}[{\includegraphics[width=1in,height=1.25in,clip,keepaspectratio]{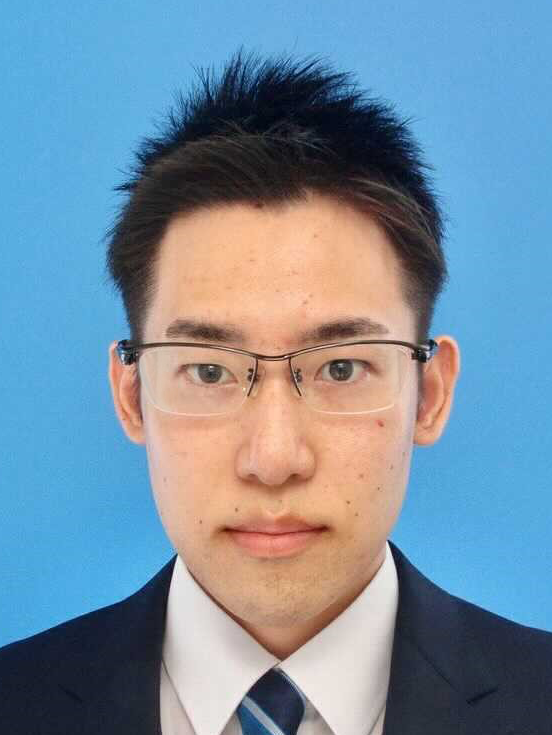}}]{Naoto Ienaga} (Member, IEEE) received his Ph.D. degree in engineering from Keio University, Yokohama, Japan, in 2020. Since 2021, he has been an Assistant Professor with the \textcolor{black}{Institute of Systems and Information Engineering} at the University of Tsukuba. He is currently working on the applications of machine learning and computer vision to solve practical problems, especially in fisheries and occupational therapy.
\end{IEEEbiography}
\newpage
\begin{IEEEbiography}[{\includegraphics[width=1in,height=1.25in,clip,keepaspectratio]{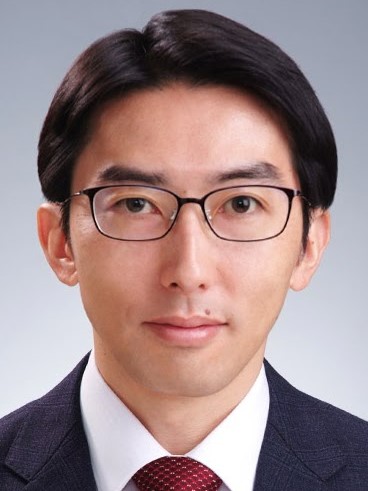}}]{Yoshihiro Kuroda}
(Member, IEEE) received his Ph.D. degree in informatics from Kyoto University, Kyoto, Japan, in 2005. He was an Assistant Professor with the Graduate School of Engineering Science at Osaka University from 2006 to 2013, an Associate Professor with the Cybermedia Center at Osaka University from 2013 to 2016, and an Associate Professor with the Graduate School of Engineering Science at Osaka University from 2016 to 2019. Since 2019, he has been a Professor with the \textcolor{black}{Institute of Systems and Information Engineering }
at the University of Tsukuba. His research interests include haptic interaction technologies and biomedical engineering.
\end{IEEEbiography}

\begin{thebibliography}{1}
\bibliographystyle{IEEEtran}
\bibitem{warmAndCool} L. A. Jones and H. Ho. ``Warm or cool, large or small? The challenge of thermal displays,'' \textit{IEEE Transactions on Haptics}, vol. 1, no. 1, pp. 53-70, 2008.
\bibitem{temperatureSensitivity} D. R. Kenshalo. ``Correlations of temperature sensitivity in man and monkey, a first approximation,'' in \textit{Collection of Sensory functions of the skin in primates}, pp. 305-330, 1976.

\bibitem{hot1} J. H. Jun et al. ``Laser-induced thermoelastic effects can evoke tactile sensations,'' \textit{Scientific reports}, vol. 5, no. 1, pp. 1-16, 2015.
\bibitem{hot2} R. A. Meyer, R. E. Walker, and V. B. Mountcastle. ``A laser stimulator for the study of cutaneous thermal and pain sensations,'' \textit{IEEE transactions on biomedical engineering}, no. 1, pp. 54-60, 1976.
\bibitem{hot3} S. Saga. ``HeatHapt thermal radiation-based haptic display,'' in \textit{Collection of Haptic Interaction}, pp. 105-107, 2015.
\bibitem{hot4} S. Saga. ``Thermal-radiation-based haptic display—laser-emission-based radiation system,'' in \textit{Proceedings of International AsiaHaptics conference}, pp. 196-197, 2018.
\bibitem{hot5} A. Sakai, T. Yamaguchi, H. Mitake, and S. Hasegawa. ``Contactless Warmth Display Using Visible Light LED for HMDVR (in Japanese),'' \textit{Transactions of the Virtual Reality Society of Japan}, vol. 24, no. 1, pp. 83-92, 2019.

\bibitem{coldAir} J. Xu, Y. Kuroda, S. Yoshimoto, and O. Oshiro. ``Non-contact cold thermal display by controlling low-temperature air flow generated with vortex tube,'' in \textit{Proceedings of 2019 IEEE World Haptics Conference (WHC)}, pp. 133-138, 2019.
\bibitem{coldAir2} J. Xu, S. Yoshimoto, N. Ienaga, and Y. Kuroda. ``Intensity-adjustable non-contact cold sensation presentation based on the vortex effect,'' \textit{IEEE Transactions on Haptics}, vol. 15, no. 3, pp. 592-602, 2022.
\bibitem{coldAir3} K. Makino, J. Xu, A. Kaneko, N. Ienaga and Y. Kuroda. ``Spatially Continuous Non-Contact Cold Sensation Presentation Based on Low-Temperature Airflows,'' \textit{Proceedings of 2023 IEEE World Haptics Conference (WHC)}, pp. 223-229, 2023.

\bibitem{peltier1} H. Ho and L. A. Jones. ``Material identification using real and simulated thermal cues,'' in \textit{Proceedings of The 26th Annual International Conference of the IEEE Engineering in Medicine and Biology Society}, vol. 1, pp. 2462-2465, 2004.
\bibitem{peltier2}A. Yamamoto, B. Cros, H. Hasgimoto, and T. Higuchi. ``Control of thermal tactile display based on prediction of contact temperature,'' in \textit{Proceedings of IEEE International Conference on Robotics and Automation}, vol. 2, pp. 1536-1541, 2004.
\bibitem{peltier3} J. Citerin, A. Pocheville, and A. Kheddar. ``A touch rendering device in a virtual environment with kinesthetic and thermal feedback,'' in \textit{Proceedings of IEEE International Conference on Robotics and Automation}, pp. 3923-3928, 2006.
\bibitem{peltier4} B. Deml, A. Mihalyi, and G. Hanning. ``Development and experimental evaluation of a thermal display,'' in \textit{Proceedings of EuroHaptics Conference}, pp. 257-262, 2006.
\bibitem{peltier5} K. Sato, and T. Maeno. ``Presentation of rapid temperature change using spatially divided hot and cold stimuli,'' \textit{Journal of Robotics and Mechatronics}, vol. 25, no. 3, pp. 497-505, 2013.
\bibitem{peltier6} R. L. Peiris, W. Peng, Z. Chen, L. Chan, and K. Minamizawa. ``Thermovr: exploring integrated thermal haptic feedback with head mounted displays,'' in \textit{Proceedings of the 2017 CHI Conference on Human Factors in Computing Systems}, pp. 5452-5456, 2017.
\bibitem{peltier7} K. Kushiyama et al. ``Thermoesthesia: about collaboration of an artist and a scientist,'' in \textit{Collection of ACM SIGGRAPH 2006 Sketches}, pp. 142-es, 2006. 
\bibitem{water1} M. Sakaguchi, K. Imai and K. Hayakawa. ``Development of high-speed thermal display using water flow,'' in \textit{Proceedings of International Conference on Human Interface and the Management of Information}, pp. 233-240, 2014.
\bibitem{water2} K. Hayakawa, K. Imai, R. Honaga, and M. Sakaguchi. ``High-speed thermal display system that synchronized with the image using water flow,'' in \textit{Collection of Haptic Interaction}, pp. 69-74, 2015.
\bibitem{fan} J. Dionisio, V. Henrich, U. Jakob, A. Rettig, and R. Ziegler. ``The virtual touch: haptic interfaces in virtual environments,'' \textit{Computer and Graphics}, vol. 21, pp. 459-468, 1997.
\bibitem{fan1} K. Ogasahara, and M. Sakaguchi. ``Influence on human body by thermal sensation stimulation using wind (in japanese),'' in \textit{Proceedings of The 24th Annual Conference of the Virtual Reality Society of Japan}, 3 pages, 2019.
\bibitem{dryIce} M. Nakajima, K. Hasegawa, Y. Makino, and H. Shinoda. ``Remotely displaying cooling sensation via ultrasound-driven air flow,'' \textit{Proceedings of 2018 IEEE Haptics Symposium (HAPTICS)}, pp. 340-343, 2018.


\bibitem{coldRadient} Y. Kume, T. Mizuno, and R. Yonezawa. ``A non-contact thermal display system by higher and lower temperature radiation sources (in Japanese),'' in \textit{Proceedings of The 27th Annual Conference of the Virtual Reality Society of Japan}, 4 pages, 2022.
\bibitem{hokoyama2017mugginess} K. Hokoyama, Y. Kuroda, G. Kato, K. Kiyokawa, and H. Takemura. ``Mugginess sensation: Exploring its principle and prototype design,'' in \textit{Proceedings of 2017 IEEE World Haptics Conference (WHC)}, pp. 563-568, 2017.
\bibitem{mist0} M. Nakajima, K. Hasegawa, Y. Makino, and H. Shinoda. ``Remotely displaying cooling sensation using ultrasound mist beam,'' in \textit{Proceedings of International AsiaHaptics conference}, pp. 85-87, 2018.
\bibitem{mist1} M. Nakajima, Y. Makino, and H. Shinoda. ``Remote cooling sensation presentation controlling mist in midair,'' in \textit{Proceedings of 2020 IEEE/SICE International Symposium on System Integration (SII)}, pp. 1238-1241, 2020.
\bibitem{mist2} M. Nakajima, K. Hasegawa, Y. Makino, and H. Shinoda. ``Spatiotemporal pinpoint cooling sensation produced by ultrasound-driven mist vaporization on skin,'' \textit{IEEE Transactions on Haptics}, vol. 14, no. 4, pp. 874-884, 2021.

\bibitem{continuousThermal1} A. Manasrah, N. Crane, R. Guldiken, and K. B. Reed. ``Perceived cooling using asymmetrically-applied hot and cold stimuli,'' \textit{IEEE Transactions on Haptics}, vol. 10, no. 1, pp. 75-83, 2016.
\bibitem{continuousThermal2} A. Manasrah, N. Crane, R. Guldiken, and K. B. Reed. ``Asymmetrically-applied hot and cold stimuli gives perception of constant heat,'' in \textit{Proceedings of 2017 IEEE World Haptics Conference (WHC)}, pp. 484-489, 2017.
\bibitem{continuousThermal3} M. Hojatmadani and K. Reed. ``Asymmetric cooling and heating perception,'' in \textit{Proceedings of International Conference on Human Haptic Sensing and Touch Enabled Computer Applications}, pp. 221-233, 2018.
\bibitem{vision} J. P. Nichol. \textit{A Cyclopaedia of the Physical Sciences}, Richard Griffin and Company, 1857.
\bibitem{thermoreceptors} I. Darian-Smith and K. O. Johnson. ``Thermal sensibility and thermoreceptors,'' \textit{Journal of Investigative Dermatology}, vol. 69, no. 1, pp. 146-153, 1977.
\bibitem{temperatureReceptors} D. C. Spray. ``Cutaneous temperature receptors,'' \textit{Annual Review of Physiology}, vol. 48, pp. 624-638, 1986.
\bibitem{thresholds} D. R. Kenshalo, E. H. Charles, and B. W. Paul. ``Warm and cool thresholds as a function of rate of stimulus temperature change,’’ \textit{Perception and Psychophysics}, vol. 3, no. 2, pp. 81-84, 1968.
\bibitem{heattransfer1} I. V. Lienhard and H. John. \textit{A heat transfer textbook}, Phlogiston Press, 2005.
\bibitem{heattransfer2} T. L. Bergman, A. S. Lavine, F. P. Incropera, and D. P. DeWitt. \textit{Introduction to heat transfer}, John Wiley and Sons, 2011.

\bibitem{vortextube} G. J. Ranque. ``Methods and apparatus for obtaining from a fluid under pressure two currents of fluids at different temperatures,'' \textit{US Patent}, No. 1952281, 1934, doi: 10.1016/j.energy.2019.116147.
\bibitem{jetEngineering} 
H. Martine. ``Heat and Mass Transfer Between Impinging Gas jets and Solid Surfaces,'' \textit{Advances in Heat Transfer}, vol.~13, pp.~1--60, 1970.
\end{thebibliography}
\end{document}